\documentclass[a4paper,american,aps,citeautoscript,floatfix,longbibliography,pdftex,pre,
superscriptaddress,showpacs,preprint]{revtex4-1}

\usepackage{adjustbox}

\usepackage{enumerate,tensor}
\usepackage{natbib} 
\usepackage{hyperref}
\usepackage{amsmath,amssymb,amsmath}
\usepackage{graphicx}
\usepackage{dcolumn}
\usepackage{bm}
\usepackage{color}

\newcommand{\ie}{i.\,e.}%

\makeatletter
\def\paragraph{\@startsection{paragraph}{4}{10pt}{-1.25ex plus -1ex minus -.1ex}{0ex plus 0ex}{\normalsize\textit}}
\renewcommand\@biblabel[1]{#1}
\renewcommand\@makefntext[1]%
{\noindent\makebox[0pt][r]{\@thefnmark\,}#1}
\DeclareRobustCommand\onlinecite{\@onlinecite}
\def\@onlinecite#1{\begingroup\let\@cite\NAT@citenum\citealp{#1}\endgroup}
\def\tagform@#1{\maketag@@@{\ignorespaces#1\unskip\@@italiccorr}}
\let\orgtheequation\theequation
\def\theequation{(\orgtheequation)}
\makeatother

\begin{document}

\title{Analysis of classical phase space and energy transfer for two rotating dipoles in an electric field}

\author{Rosario Gonz\'alez-F\'erez}
\affiliation{Instituto Carlos I de F\'{\i}sica Te\'orica y Computacional,
and Departamento de F\'{\i}sica At\'omica, Molecular y Nuclear,
  Universidad de Granada, 18071 Granada, Spain} 

\author{Manuel I\~narrea}
\affiliation{\'Area de F\'{\i}sica, Universidad de La Rioja, 26006 Logro\~no, La Rioja, Spain}

  \author{J. Pablo Salas}
\affiliation{\'Area de F\'{\i}sica, Universidad de La Rioja, 26006 Logro\~no, La Rioja, Spain}

\author{Peter Schmelcher}
\affiliation{The Hamburg Center for Ultrafast Imaging, Luruper Chaussee 149, 22761 Hamburg, Germany}
\affiliation{Zentrum f\"ur Optische Quantentechnologien, Universit\"at
  Hamburg, Luruper Chaussee 149, 22761 Hamburg, Germany} 
  
\date{\today}
\begin{abstract} 
We explore the classical dynamics of two interacting rotating dipoles that are fixed in the space and
exposed to an external homogeneous electric field.
Kinetic energy transfer mechanisms between the dipoles are investigated
varying both the amount of initial excess kinetic energy of one of them and the strength of the electric field.
In the field-free case, and depending on the initial excess energy an abrupt transition
between equipartition and non-equipartition regimes is encountered. The study of the
phase space structure of the system as well as the formulation of the Hamiltonian in an appropriate
coordinate frame provide a thorough understanding of this sharp transition.
When the electric field is turned  on, the kinetic energy transfer mechanism is significantly more complex
and the system goes through different regimes of equipartition and non-equipartition of the energy 
including chaotic behavior.
\end{abstract}
\maketitle

\section{Introduction}
\label{sec:introduction}

The mechanism of energy exchange between molecules, 
mediated either by the Coulomb, dipole-dipole or van-der-Waals interactions
is an active research area with several intriguing perspectives in physics, chemistry, biology and material sciences. 
The wide range of applications cover, for instance, the photosynthesis of plants and 
bacteria~\cite{van,engel,fotosintesis,wu,Burghard}, 
 the emission of light of  organic materials~\cite{Saikin,Melnikau,Qiao}, and  molecular crystals~\cite{Davydov,Silbey,Wright}. On the other hand side, cooling and trapping cold molecules in an optical lattice
 allows to fix their positions while exploiting their interactions~\cite{krems,weidemuller}. The latter becomes particularly
 interesting for strongly polar diatomic systems where the dipole--dipole interaction is
 sufficiently long--range that novel structural as well as dynamical and collective
 behaviors can be expected~\cite{lahaye,rey,Lewenstein}.
 External electric fields provide then a versatile tool to control these interactions, e.g. the alignment of the
 dipoles with the field~\cite{krems2}. 

One of the most popular approaches to investigate the energy transfer in a many--body system is to describe it by 
a linear chain of nonlinear oscillators with different coupling between them.
These models are based on the seminal work of  Fermi, Pasta and Ulam~\cite{fpu}, the so-called FPU system. 
This work was the first to realize that, in the infinite time limit, this system of nonlinear oscillators  does 
not reach the expected 
smooth  energy-equipartition behavior. 
After several decades of research and a plethora of works, see for instance 
Ref.~\cite{Ford,Berman,Dauxois,Gallavotti,Mussot,Bambusi,penati}, the 
question concerning the energy sharing mechanism  in a chain of nonlinear oscillators, and, therefore, in 
a many--body system, can be considered  still an open question.
Furthermore, in references~\cite{Ratner1,Ratner2,Ratner3} the energy flow in a
linear chain of interacting rotating dipoles and in a two--dipole system are explored.
For the two-dipole system, the authors conclude the existence of a critical excitation energy up to
which there is no energy transfer.

In order to provide further insights in the energy transfer mechanisms in dipole chains, 
in this work, we consider two interacting rotating dipoles exposed to an external electric field. 
The aim is to investigate the classical phase space in relation to
the energy transfer mechanism between the two rotors. 
Assuming that their positions are fixed in space, we employ a classical description  of their
internal dynamics within the rigid rotor approximation. 
A certain amount of kinetic energy is then given
to one of the dipoles and the energy transfer mechanism between the two dipoles is
explored as the excess kinetic energy and the field strength are varied.
For the field-free system, we encounter 
energy-equipartition and non-equipartition regimes  depending on the 
initial excess energy.  
In the field-dressed system, there exists a competition between the anisotropic
dipole-dipole interaction of the rotors and
the electric field interaction.
If the strengths of these two interactions are comparable, the classical dynamics is chaotic.
As the strength of the electric field increases, and the field interaction dominates, we encounter
an energy-equipartition regime, that is followed by a energy-localized one for even stronger fields.

The paper is organized as follows: In \autoref{sec:hamiltonian} we establish the classical rotational 
Hamiltonian governing the dynamics of  two identical  rotating dipoles in an external
electric field with fixed spatial 
positions. The equations of motion and the critical points in an invariant manifold are also presented. 
\autoref{sec:en_tra_no_dc_field} and~\autoref{sec:en_tra_dc_field} are devoted to the
investigation of the exchange  of energy  between the two rotors in the field-free case 
and in the presence of the external field,
respectively. The conclusions are provided in \autoref{sec:conclusions}.

\section{Classical Hamiltonian and equations of motions}
\label{sec:hamiltonian}

We consider  two identical dipoles, fixed in space and separated by a distance $a_l$ along
the Laboratory Fixed Frame (LFF) $X$-axis. 
Here, we employ the rigid rotor approximation to describe the dynamics
of the two dipoles.
In the presence of an  external  homogeneous time-dependent electric field parallel
to the LFF $Z$-axis and with strength ${\cal E}_s(t)$, the interaction potential,
${\cal V}\equiv{\cal V}(\theta_1, \phi_1,\theta_2, \phi_2,t)$, 
is given by
\begin{widetext}
\begin{equation}
\label{eq:potential}
{\cal V}=-\mu {\cal E}_s(t)(\cos\theta_1 +\cos\theta_2) + \frac{\mu^2}{4\pi\epsilon_0 a_l^3}
[\cos\theta_1\cos\theta_2
+\sin\theta_1\sin\theta_2(\sin\phi_1\sin\phi_2-2\cos\phi_1\cos\phi_2)],
\end{equation}
\end{widetext}
where $(\theta_i, \phi_i)$, with $i=1,2$,  represent the Euler angles of each rotor. 
The first term in~\autoref{eq:potential} stands for the interaction of the dipole moment, $\mu$, of the two rotors 
with the external electric field of strength ${\cal E}_s(t)=E_sf(t)$ that is turned on with the linear function
\begin{equation}
\label{pulse}
f(t)=\left \{
      \begin{matrix} 
         \displaystyle  \frac{t}{t_1} & \mbox{if} & 0 \le t < t_1\\[2ex]
         1 & \mbox{if} & t \ge t_1\, .
        \end{matrix}
   \right .
   \end{equation}
The last term in~\autoref{eq:potential}  represents the dipole-dipole interaction between the  two rotors.
The classical Hamiltonian describing the rotational motion  of this system reads
\begin{equation}
\label{eq:ham2}
{\cal H} =\sum_{i=1}^2 \frac{1}{2 I}  \bigg[ P_{\theta_i}^2+
 \frac{P_{\phi_i}^2}{\sin^2\theta_i}\bigg]+{\cal V}, 
\end{equation}
where $I$ is the moment of inertia of the dipoles, and
where the first two terms stand for the rotational kinetic energy of the dipoles.
Expression \ref{eq:ham2} defines a $4$ degree-of-freedom Hamiltonian dynamical system in 
$(\theta_1, \phi_1, \theta_2, \phi_2)$ and in the  corresponding momenta
$( P_{\theta_1}, P_{\phi_1}, P_{\theta_2}, P_{\phi_2})$.
For the sake of simplicity, it is convenient to handle a dimensionless Hamiltonian. To this end, we
express energy in units of the molecular rotational constant $B=\hbar^2/2 I$ and time in units of the characteristic
time $t_B=\hbar/2B$. In this way, we arrive at the dimensionless Hamiltonian given by
\begin{equation}
\label{eq:ham33}
 H \equiv  \frac{{\cal H}}{B}= \sum_{i=1}^2 \bigg[ P_{\theta_i}^2+
\frac{P_{\phi_i}^2}{\sin^2\theta_i}\bigg] + V,
\end{equation}
with the rescaled potential, $V\equiv V(\theta_1, \phi_1,\theta_2, \phi_2,t)$, 
being
\begin{widetext}
\begin{equation}
V=-f(t) \beta(\cos\theta_1+\cos\theta_2)+
\chi
[\cos\theta_1\cos\theta_2 
+\sin\theta_1\sin\theta_2(\sin\phi_1\sin\phi_2-2\cos\phi_1\cos\phi_2)], 
\end{equation}
\end{widetext}
where the dimensionless parameters 
\begin{equation}
\label{chi}
\chi=\frac{\mu^2}{4 \pi\epsilon_0 a_l^3 \ B},  \qquad \text{and}\qquad  \beta=\frac{\mu E_s}{B}
\end{equation}
control the dipole-dipole and electric field interactions, respectively.

Since the two rotors are identical, the Hamiltonian~\eqref{eq:ham33} posseses an  
exchange symmetry of even character, and it presents the following invariant manifold
\begin{equation*}
\begin{array}{l}
{\cal M} =\lbrace(\theta_1, \theta_2, P_{\theta_1}, P_{\theta_2}) \ | \ \phi_1=\phi_2= P_{\phi_1}=P_{\phi_2}=0\rbrace,\end{array}
\end{equation*}
where the dynamics  is limited to planar motions confined in the $XZ$  plane.
 In this  invariant manifold ${\cal M}$, the Hamiltonian reads
\begin{equation}
\label{eq:ham4}
H_{\cal M}\equiv {\cal E}= P_1^2+P_2^2 + V_{\cal M}(\theta_1,\theta_2,t),
\end{equation}
where $ V_{\cal M}(\theta_1,\theta_2,t)\equiv V(\theta_1,0,\theta_2,0,t)$ is the potential energy surface of this system in 
${\cal M}$.
In the rest of the paper, we focus our study on the manifold ${\cal M}$.
The Hamiltonian equations of motion arising from $H_{\cal M}$ read as follows
\begin{eqnarray}
\label{ecumoviA}
\dot \theta_1 &=& 2 P_1,  
\qquad \dot \theta_2 = 2 P_2, \nonumber \\[2ex] 
\dot P_1 &=& (\chi \cos \theta_2-\beta f(t)) \sin \theta_1 +2 \chi \cos \theta_1 \sin \theta_2, \\[2ex]
\dot P_2 &=& (\chi \cos \theta_1-\beta f(t)) \sin \theta_2 +2 \chi \cos \theta_2 \sin \theta_1. \nonumber
\end{eqnarray}

\subsection{The critical points of the energy surface}

\begin{table*}
\caption{Conditions of existence, stability and energy of the critical points of
$V_{\cal M}(\theta_1,\theta_2, t>t_1)$. The
saddle points are denoted by S.P.}
\centering
\begin{adjustbox}{max width=\textwidth}
\begin{tabular}{*{4}{lcll}|}
\hline\noalign{\smallskip}
Equilibrium & Existence & Stability & Energy ${\cal E}$  \\[2ex]
\noalign{\smallskip}\hline\noalign{\smallskip}
$C_1 = (\pm \cos^{-1}\beta/3\chi, \pm \cos^{-1}\beta/3\chi)$ & $\beta \le 3\chi$ & Minima &
${\cal E}_1 = -(6 \chi^2+\beta^2)/3\chi$ \\[2ex]
$C_2 = (\pm \pi, 0), \ (0, \pm \pi)$ & Always & S.P. & ${\cal E}_2 = -\chi$ \\[2ex]
$C_3=(0, 0)$ & Always & If $\beta < 3\chi$: S.P.; if $\beta > 3\chi$: Minimum &${\cal E}_3 =  \chi - 2 \beta$\\[2ex]
$C_4 = (\pm \pi, \pm \pi) \equiv (\pm \pi, \mp \pi)$ & Always &
If $\beta < \chi$: S.P.; if $\beta > \chi$: Maxima & ${\cal E}_4 =  \chi + 2 \beta$\\[2ex]
$C_5 = (\pm\cos^{-1}(-\beta/\chi), \mp\cos^{-1}(-\beta/\chi))$ & $\beta \le \chi$ & Maxima &
${\cal E}_5 = (2 \chi^2+\beta^2)/\chi$ \\[2ex]
\noalign{\smallskip}\hline
\end{tabular}
\end{adjustbox}
\label{ta:tabla1}

\end{table*}

\begin{figure}[t]
\includegraphics[scale=0.4]{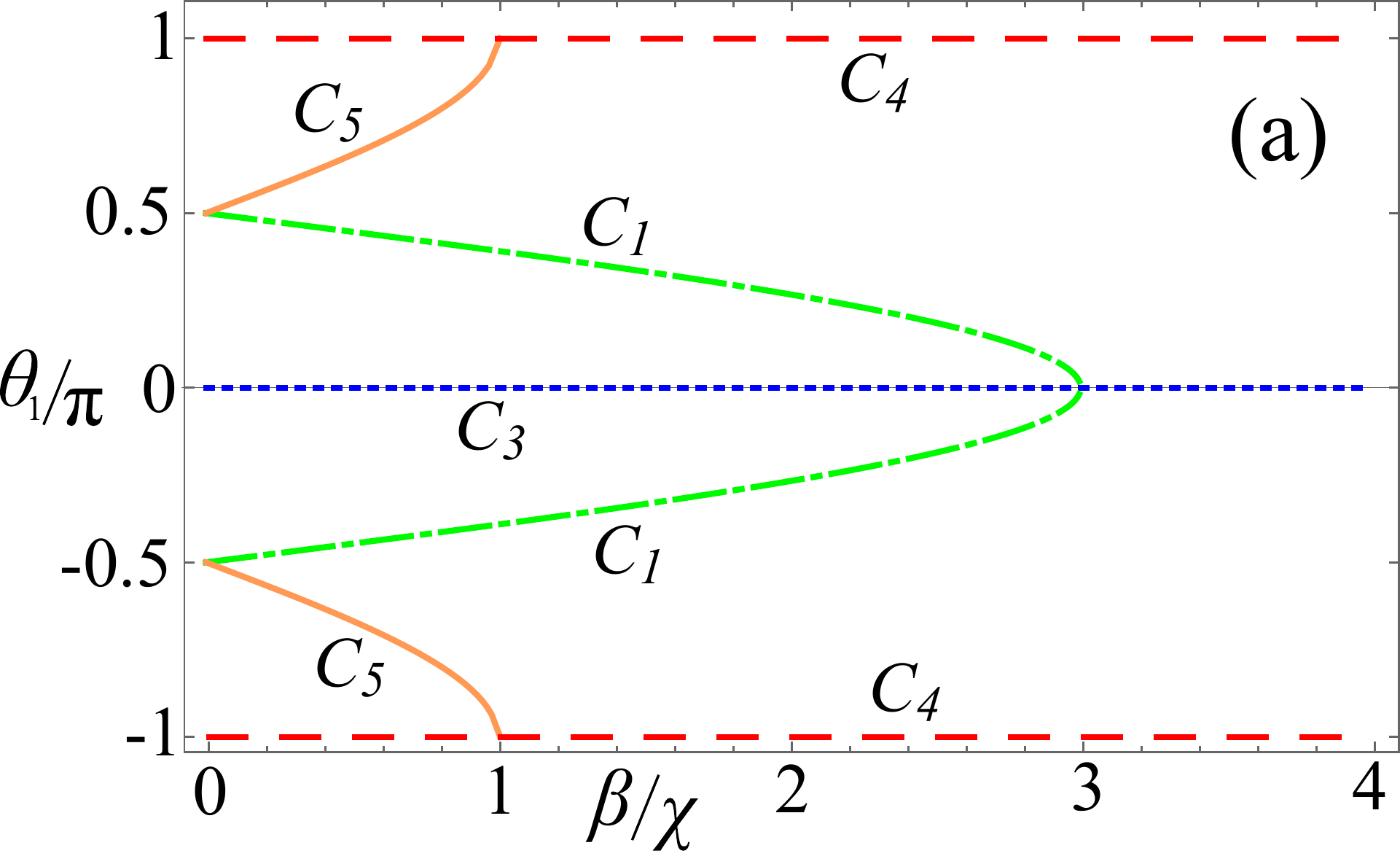}
\includegraphics[scale=0.4]{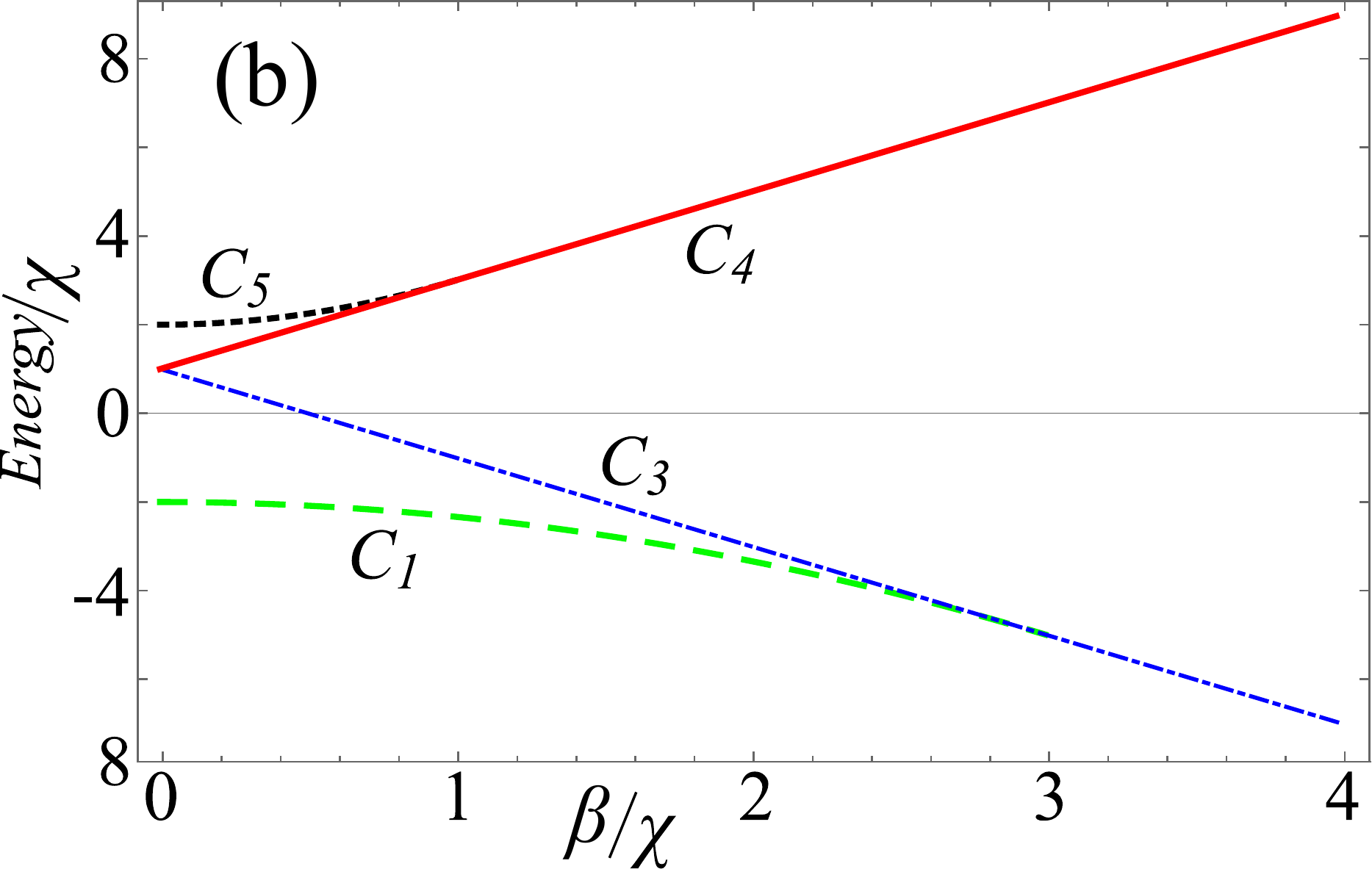}
\caption{
Evolution of  the (a) position and  (b) energy   
of the critical points $V_{\cal M}(\theta_1,\theta_2,t\le t_1)$ 
as a function of  the ratio between the  electric field parameter
$\beta$ and the dipole-dipole interaction parameter $\chi$.}
\label{fi:criticos} 
\end{figure}

Part of the dynamics can be inferred from the landscape of the potential
energy surface $V_{\cal M}(\theta_1,\theta_2,t)$, and 
its critical points, which are the
equilibrium points of the Hamiltonian flux~\eqref{ecumoviA} equated to zero.
Since the  Hamiltonian~\eqref{eq:ham4} is an even function with exchange symmetry, the critical
points are located along the directions $\theta_1=\theta_2$ and $\theta_1=-\theta_2$. 
Note that for the sake of completeness,  the polar angles $(\theta_1, \theta_2)$ are varied in the interval $[-\pi, \pi]$.
For $t\ge t_1$ ($f(t)=1$) the electric field parameter has reached it maximal value  $\beta$ and the critical
points of $V_{\cal M}(\theta_1,\theta_2,t)$ are the roots of the equations
\begin{eqnarray}
\label{critical2}
(\beta  -3 \chi \cos \theta_1) \sin \theta_1&=&0,
\quad \mbox{with} \quad \theta_2=\theta_1  \nonumber\\[-2ex]
&&\\
(\beta  + \chi \cos \theta_1) \sin \theta_1&=&0
\quad \mbox{ with} \quad \theta_2=-\theta_1.\nonumber
\end{eqnarray}
There exist  five critical points, their  conditions of existence and stability and energies 
are summarized in Table \ref{ta:tabla1}. 
The positions and  energies of these critical points as the electric field  parameter
$\beta$ increases are presented in~\autoref{fi:criticos}. 

\begin{figure*}
\includegraphics[scale=0.28]{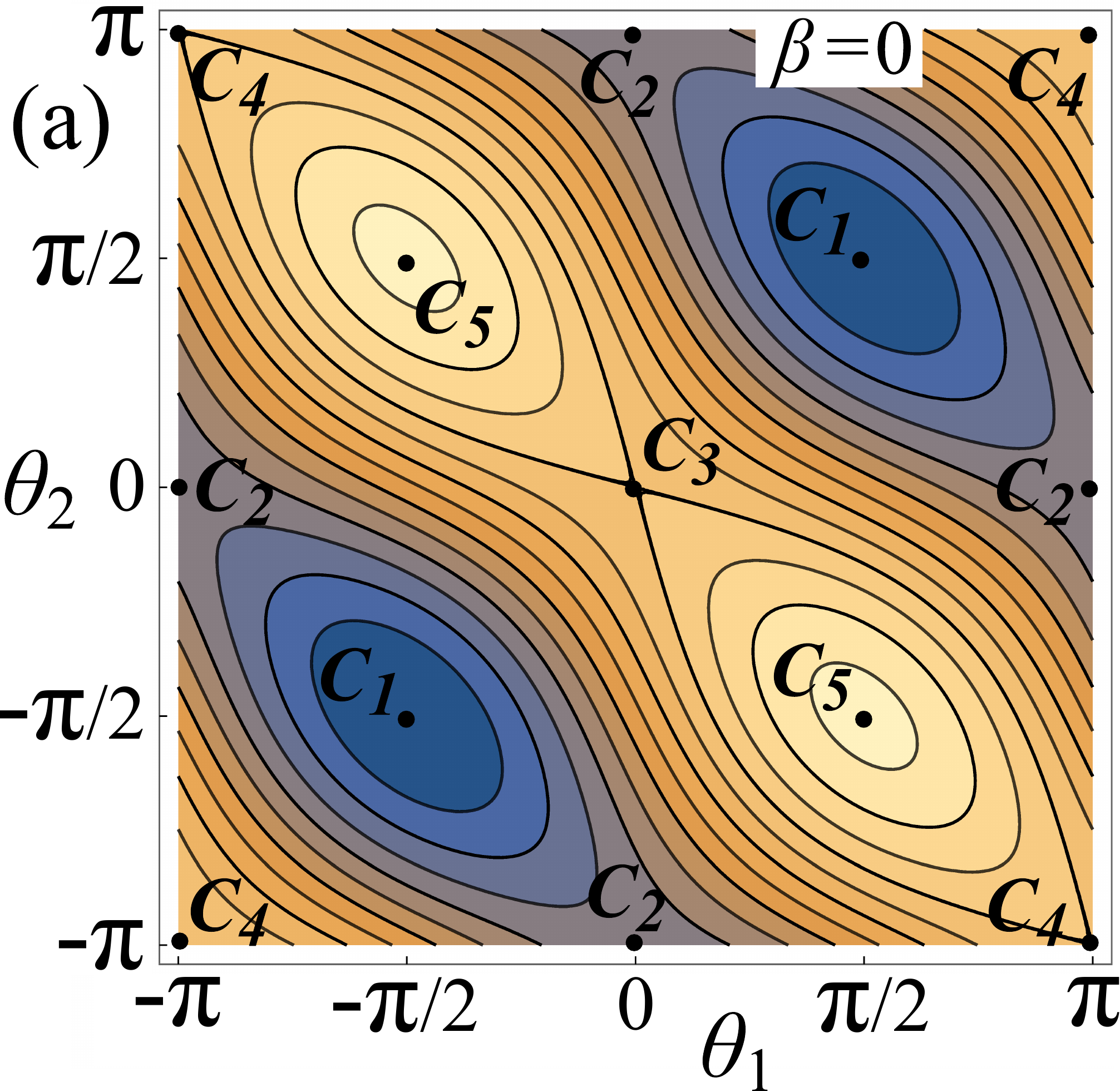} \includegraphics[scale=0.28]{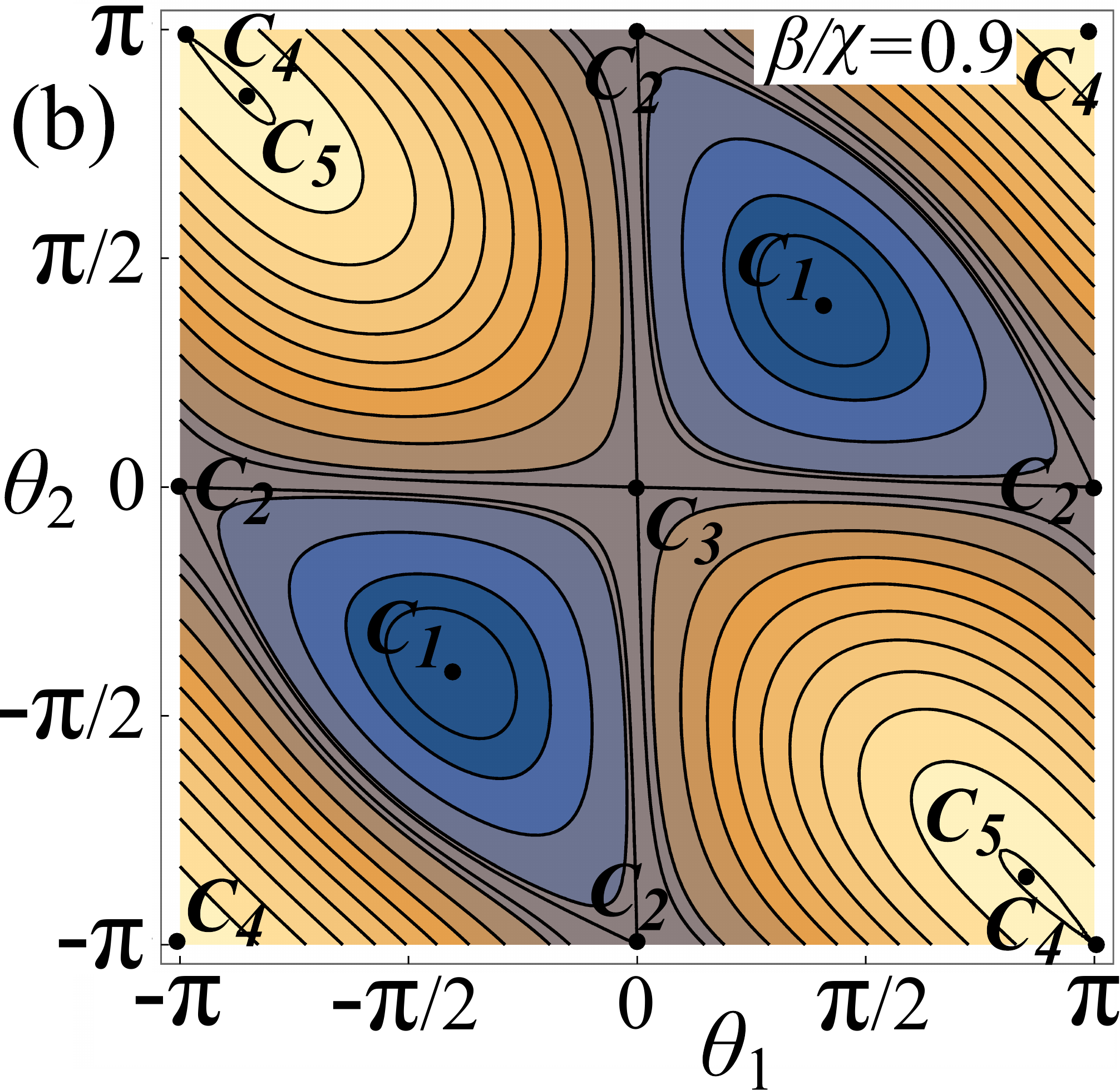}
\includegraphics[scale=0.28]{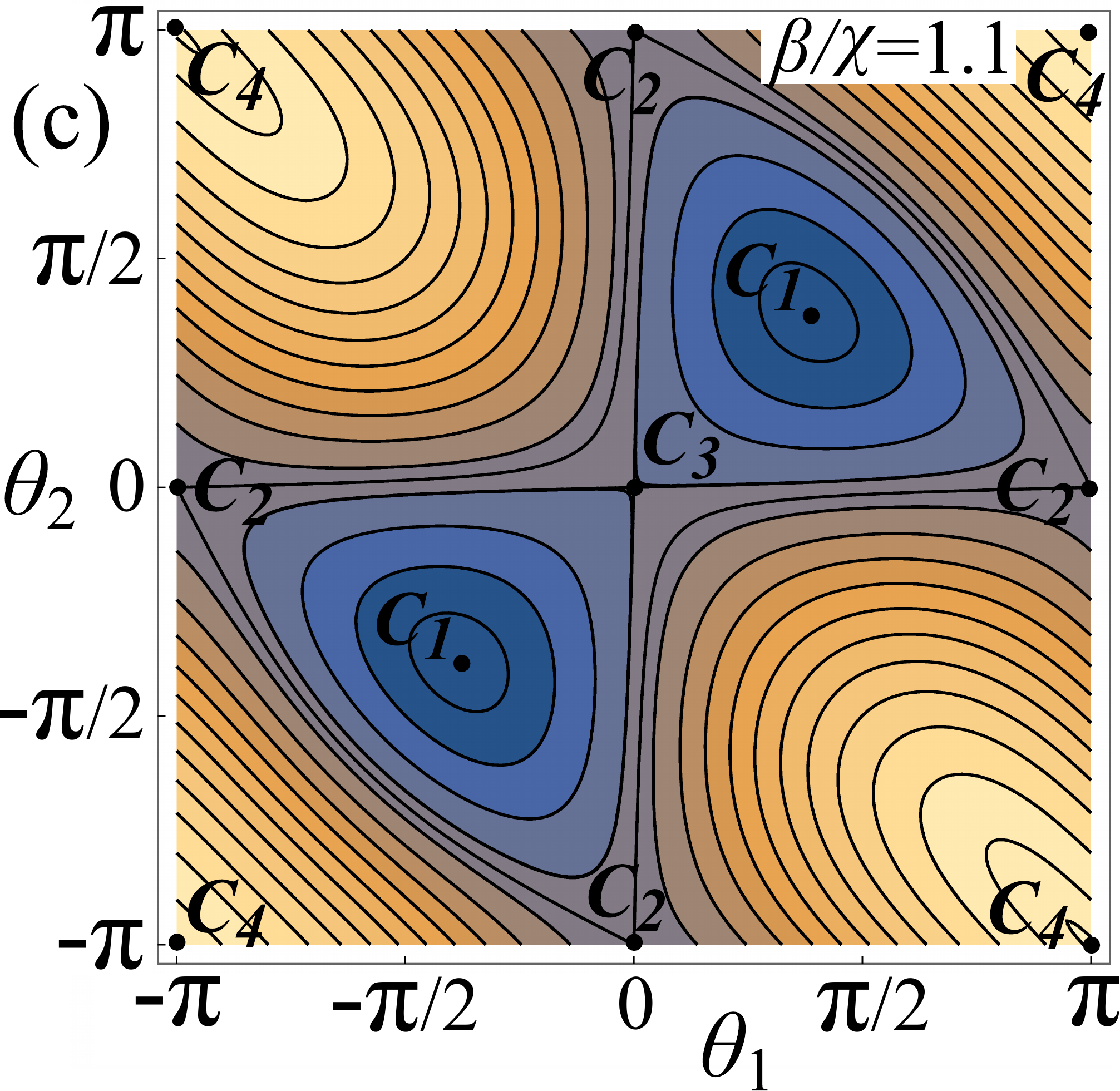} \includegraphics[scale=0.28]{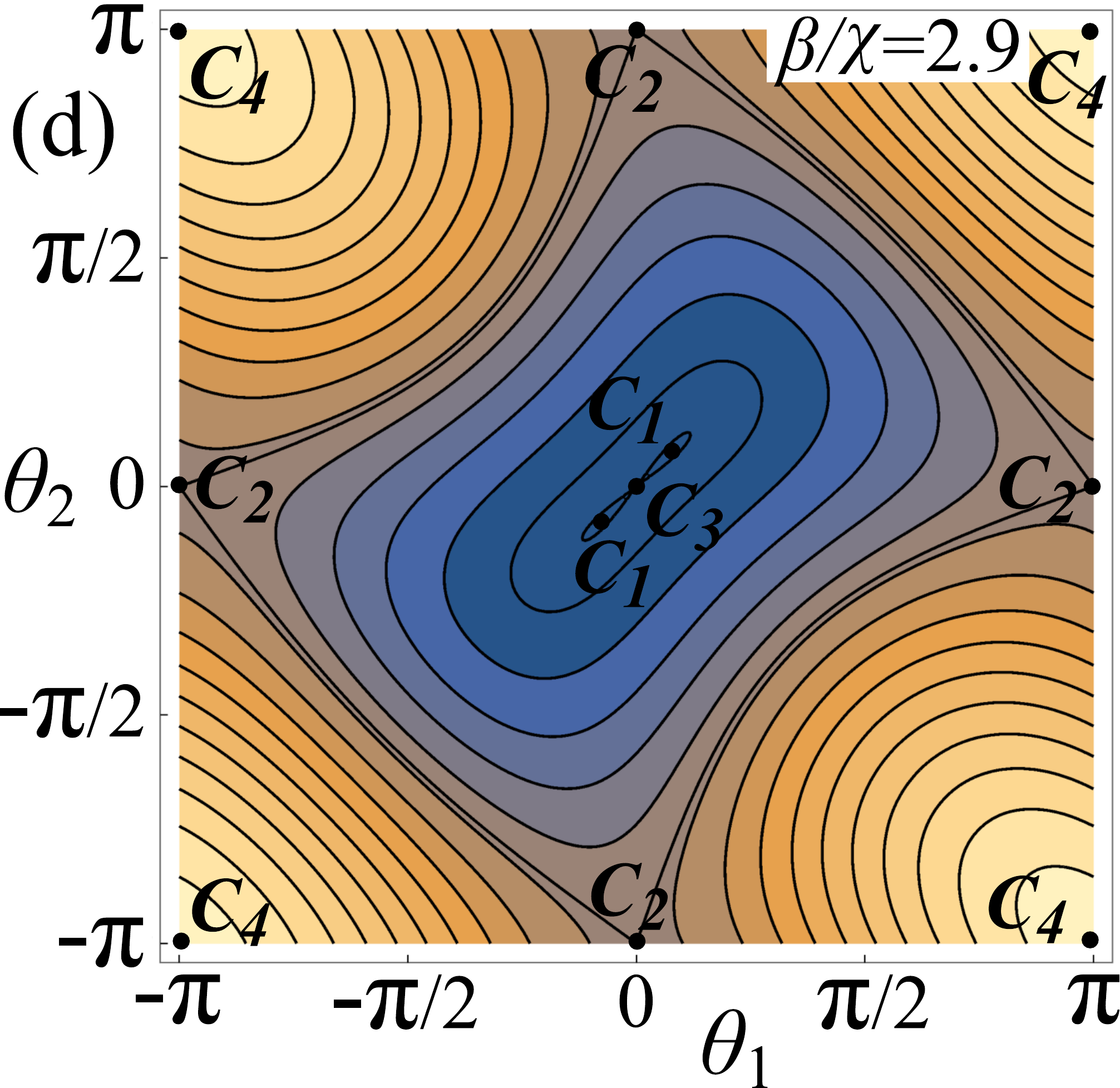}
\includegraphics[scale=0.28]{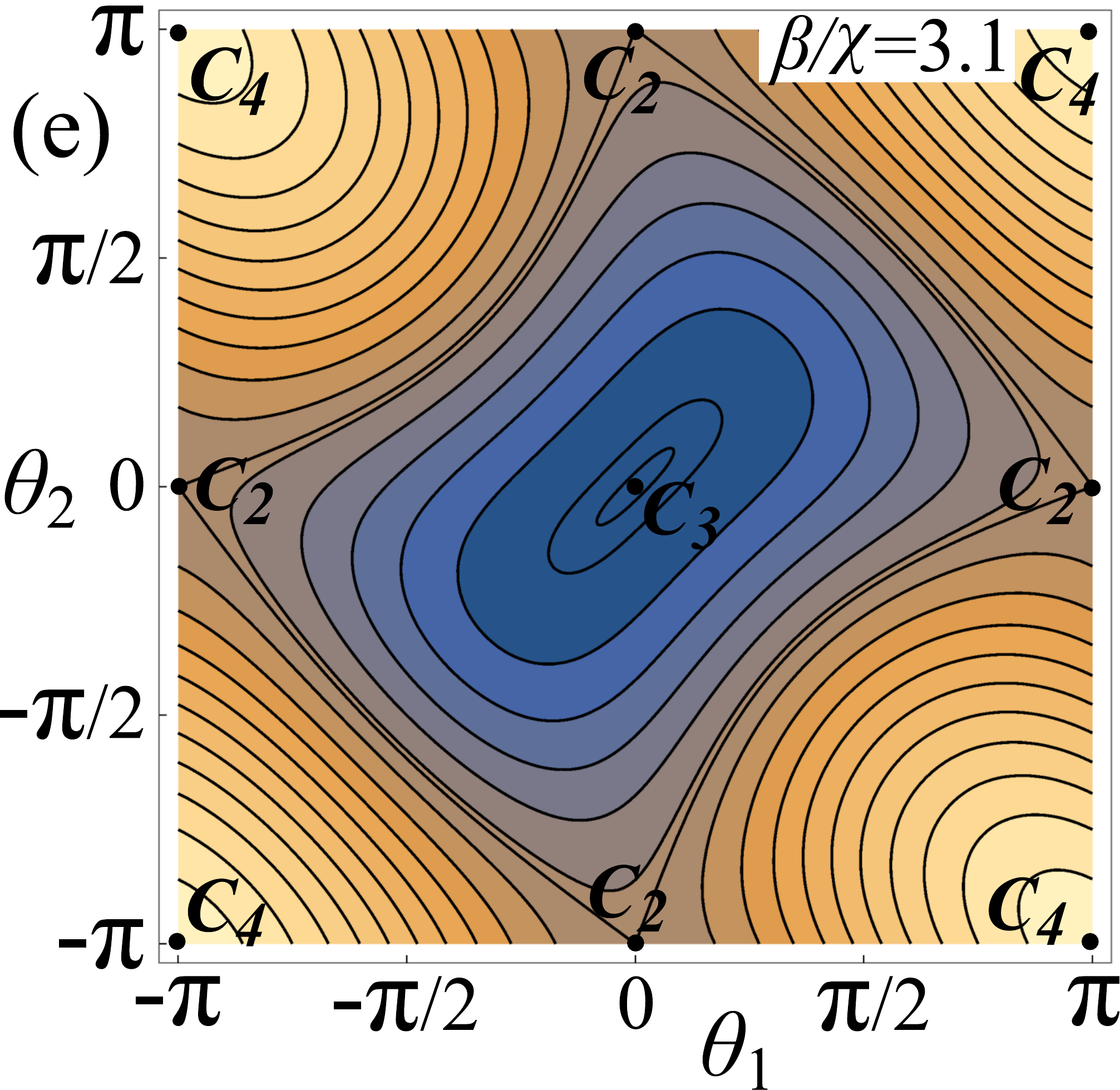} \includegraphics[scale=0.28]{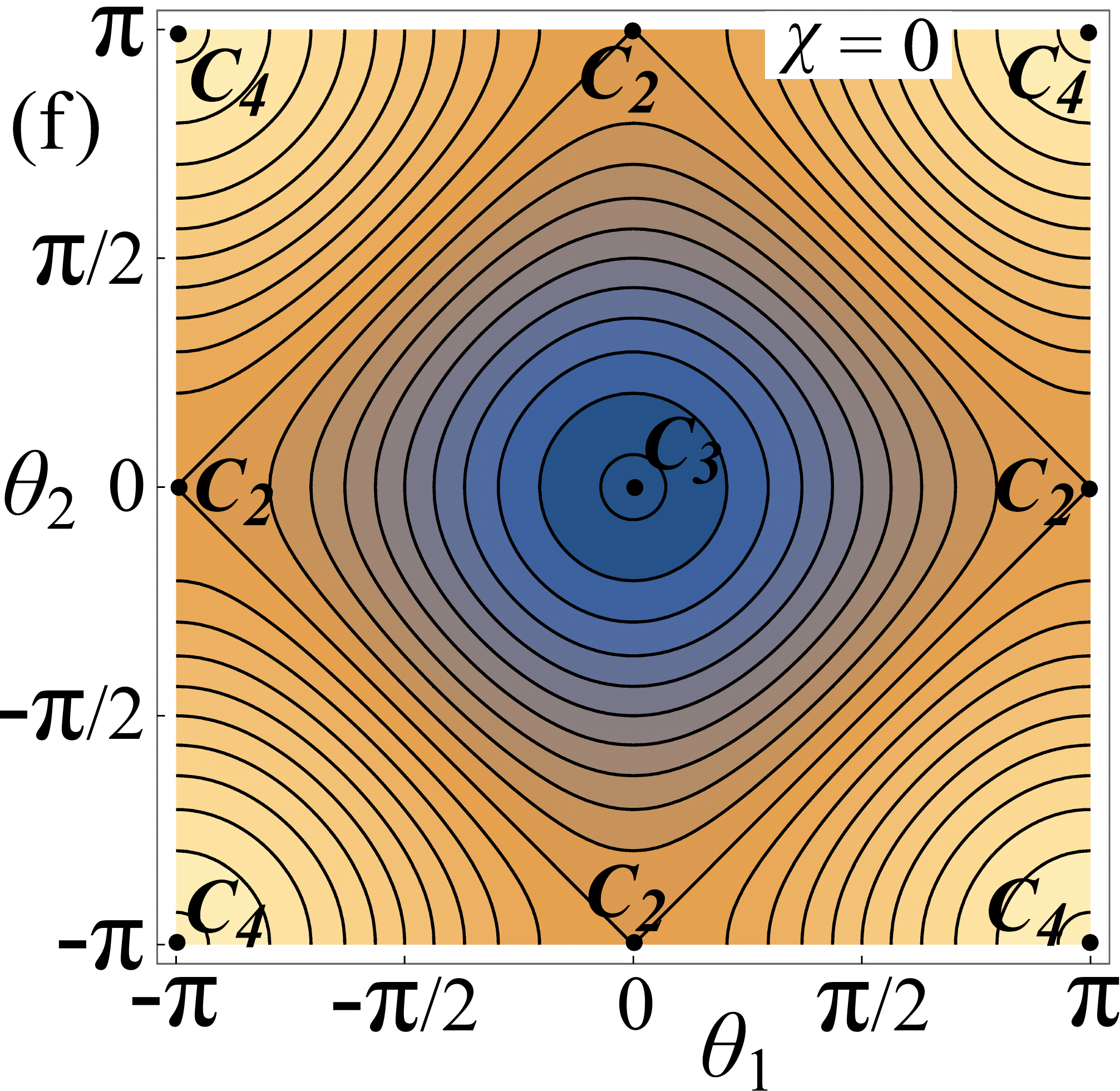}
\caption{Evolution of the landscape of the potential energy surface $V_{\cal M}(\theta_1,\theta_2,t)$ for $t>t_1$ and
 different values of the ratio between the electric field parameter and the dipole-dipole interaction $\beta/\chi$.}
\label{fi:c0}
\end{figure*}

For $0 \le \beta/\chi < 1$, the five equilibria exist. In  the field-free case $\beta=0$, the
energy surface
$V_{\cal M} (\theta_1,\theta_2,t\ge t_1)$ shows the characteristic landscape of the dipole-dipole interaction
shown in~\autoref{fi:c0}a. The minima
${\cal C}_1$ correspond to the stable {\sl head-tail} configurations of the dipoles, while the maxima ${\cal C}_5$
correspond to the unstable {\sl head-head} or {\sl tail-tail} configurations. 
Thus, if the energy of the system is below  the energy
of the saddle points ${\cal C}_2$, ${\cal E}_2=-\chi$, the two
dipoles are confined in the potential wells created by the minima ${\cal C}_1$ and they
 oscillate around the stable head-tail
 configuration. If the energy of the system is larger than ${\cal E}_2=-\chi$, and smaller than the  energy
 of the saddle points ${\cal C}_{3}$ and ${\cal C}_{4}$,  
 ${\cal E}_{3}={\cal E}_{4}=\chi$, the oscillations of the dipoles are of large amplitude but still around
 the stable head-tail configurations.
 Finally, if the total energy is larger than ${\cal E}_{3}={\cal E}_{4}=\chi$, the rotors can perform 
 complete rotations.

For  $0 < \beta/\chi < 1$,  as the ratio $\beta/\chi$ approaches to $1$,
the minima  ${\cal C}_1$ (maxima $C_5$) move towards the saddle point $C_3$ ($C_4$). 
However, the shape of the energy surface
$V_{\cal M}(\theta_1,\theta_2,t\ge t_1)$ remains qualitatively the same,
though being somewhat distorted as compared to the field free case, see~\autoref{fi:c0}b for $\beta/\chi=0.9$.
As a rough approximation, for $0 < \beta/\chi < 1$, 
the interaction due to the electric field could be 
considered as a perturbation
to the dipole-dipole interaction, which dominates the dynamics. 
For $\beta/\chi=1$, a pitchfork bifurcation takes place between
the saddle points $C_4$ and the maxima $C_5$, see~\autoref{fi:criticos}, 
and from there on only the saddle points $C_4$, which become maxima, survive, which
is illustrated in~\autoref{fi:c0}c for $\beta/\chi=1.1$.
As the electric field parameter increases in the interval $1 \le \beta/\chi < 3$, the
minima $C_1$ keep moving towards $C_3$,
see Fig. \ref{fi:criticos} and the contour plot in Fig. \ref{fi:c0}d for $\beta/\chi=2.9$.
At $\beta/\chi=3$, $C_1$ and $C_3$ collide and a second pitchfork bifurcation occurs.
From this bifurcation on, only the critical point $C_3$ survives now as  minimum, see~\autoref{fi:c0}e for $\beta/\chi=3.1$.
For $\beta/\chi \ge 3$, the shape
of the energy surface $V_{\cal M}(\theta_1,\theta_2, t=t_1)$ is qualitatively similar to 
the  $\chi=0$ case, where only the interaction due to the electric field is taken into account, 
cf.~\autoref{fi:c0}e and~\autoref{fi:c0}f.  Indeed, for  $\beta/\chi \ge 3$, 
the dipole-dipole interaction could  be considered as a perturbation to the electric field interaction.

\subsection{The rotated reference system}

\begin{figure}
\includegraphics[scale=0.3]{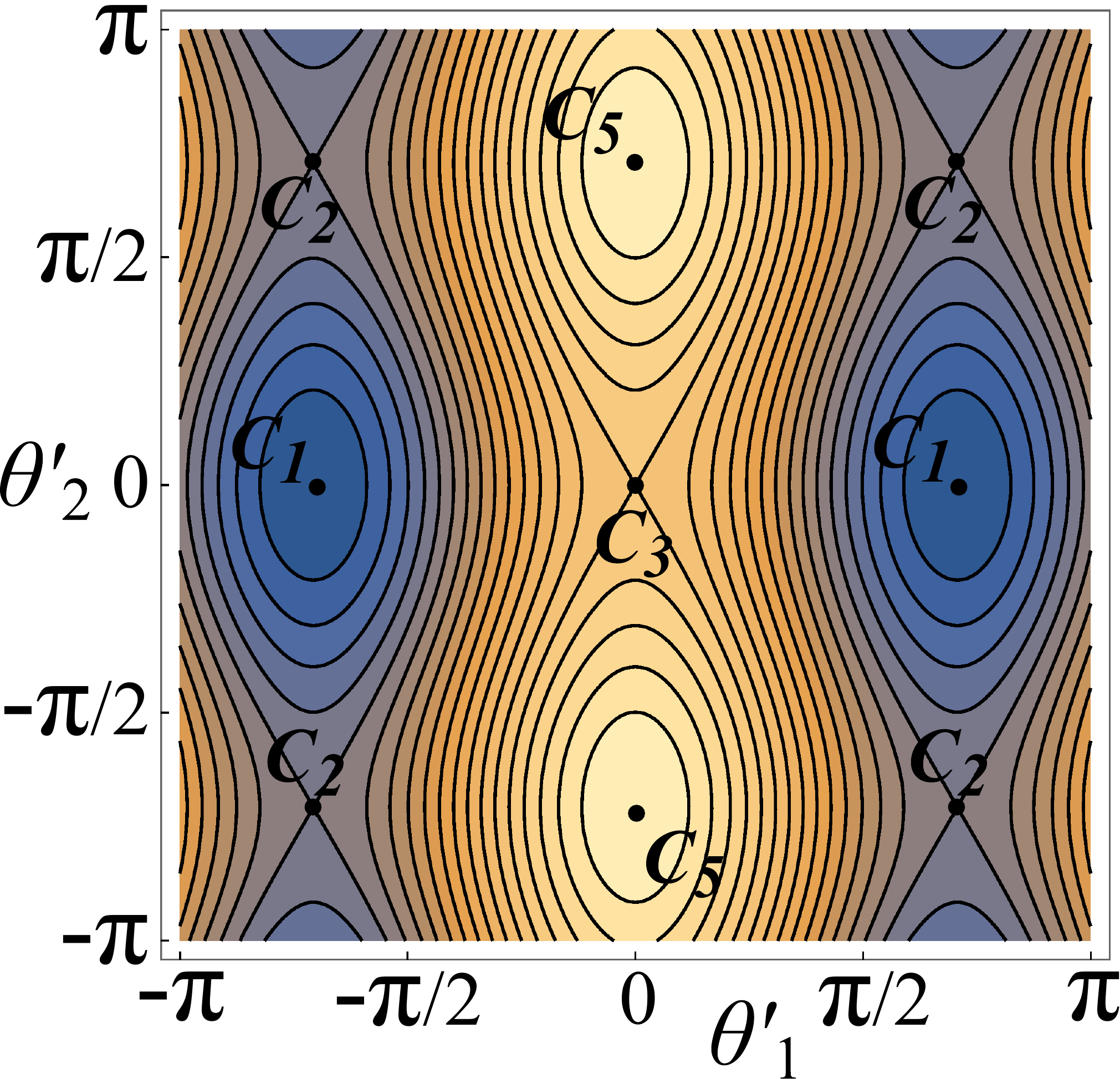}
\caption{Landscape of the rotated potential energy surface $V'(\theta_1,\theta_2,t)$
 for $\beta=0$ and $t>0$.}
\label{fi:c1}
\end{figure}  
 A  $3\pi/4$  rotation around the axis perpendicular to the plane $(\theta_1,\theta_2)$
 of the Hamiltonian~\eqref{eq:ham4} 
 takes the  equilibria along the bisector $\theta_2=\theta_1$ 
 to the axis $\theta_2=0$. This rotation is a canonical transformation between the  coordinates
 $(\theta_1, \theta_2, P_1, P_2)$ and the new ones  $(\theta'_1, \theta'_2, P'_1, P'_2)$
 given by
 \begin{eqnarray}
\label{rotation}
\theta'_1 &=& \frac{\theta_1+\theta_2}{\sqrt{2}},  \qquad 
\theta'_2 = \frac{\theta_2-\theta_1}{\sqrt{2}}, \\
P'_1&= & \frac{P_1+P_2}{\sqrt{2}},\qquad   
P'_2 = \frac{P_2-P_1}{\sqrt{2}},\nonumber
\end{eqnarray}
\noindent
and with generating function ${\cal W}$
\begin{equation}
{\cal W} = P'_1\left(\frac{\theta_1+ \theta_2}{\sqrt{2}}\right) + P'_2 \left(\frac{\theta_2-\theta_1}{\sqrt{2}}\right).
\end{equation}
\noindent
The rotated Hamiltonian $H'$ reads
\begin{equation}
\label{eq:rotated}
{ H}'=E'=P'^2_1+P'^2_2+ V'_{\cal M}(\theta'_1, \theta'_2,t),
\end{equation}
\noindent
where 
\begin{eqnarray}
\label{potenRota}
 V'_{\cal M}(\theta'_1, \theta'_2,t)&=& V'_{1}(\theta'_1)+ V'_{2}(\theta'_1) 
 \nonumber\\ 
&-& 2 \beta f(t) \cos\left(\frac{\theta'_1}{\sqrt{2}}\right) \cos\left(\frac{\theta'_2}{\sqrt{2}}\right), 
\end{eqnarray}
with 
\[
V'_{1}(\theta'_1)=\frac{3}{2} \chi \cos(\sqrt{2} \ \theta'_1),  
\quad    
V'_{2}(\theta'_2)=- \frac{1}{2} \chi \cos(\sqrt{2} \ \theta'_2).
\]
The potential $V'_{\cal M}(\theta'_1, \theta'_2,t)$ represents the potential energy of two pendula 
coupled by the external electric field. 

In the field-free case, $\beta=0$, 
the dipole-dipole Hamiltonian  is separable, ${ H}' ={ H}'_1 + { H}'_2$ with, 
 \begin{eqnarray}
\label{rotation2}
{ H}'_1 &=& E_1'= P'^2_1 + \frac{3}{2} \chi \cos(\sqrt{2} \ \theta'_1),\nonumber \\[2ex] 
{ H}'_2 &=& E_2'= P'^2_2 - \frac{1}{2} \chi \cos(\sqrt{2} \ \theta'_2),
\end{eqnarray}
and the dynamics is that of 
two uncoupled pendula. The contour plot of $ V'_{\cal M}(\theta'_1, \theta'_2,t)$ for $\beta=0$
and $t>t_1$ 
is depicted in~\autoref{fi:c1}.

\section{Energy Transfer in the field-free case}
\label{sec:en_tra_no_dc_field}

In this section, we explore the energy  transfer mechanism between 
the two field-free rotors assuming that, initially, they do not have the same kinetic energy. 
Indeed,  we assume that initially the two rotors are
at rest, with zero kinetic energy, in the bottom of the potential well ${\cal C}_1$, \ie, $\theta_1=\theta_2=\pi/2$, 
in the stable head-tail
configuration with  total energy $-2 \chi$. 
Then, a certain amount of kinetic energy $\delta {\cal K}$ is given to the first dipole, in such a way that 
 the initial conditions at $t=0$ are 
\begin{equation}
\label{ci}
\theta_1(0)=\theta_2(0)=\frac{\pi}{2}, \quad P_1(0)=\sqrt{\delta {\cal K}},\quad P_2(0)=0.
\end{equation}
With these initial conditions, the Hamiltonian equations of motion~\eqref{ecumoviA} for $\beta=0$ are integrated 
up to a final time $t_f$.
During the numerical integration, we compute the normalized time average  of the kinetic energy
of each dipole, $\widehat{P_i^2}$,  given by
\begin{equation}
\label{average}
\widehat{P_i^2}=\frac{\langle P_i^2 \rangle}{\langle P_1^2 \rangle + \langle P_2^2 \rangle},\quad
\langle P_i^2 \rangle = \frac{1}{t_f-t_1} \int_{t_1}^{t_f} P_i^2(t) dt .
\end{equation}
Note that in the field-free case $\beta=0$, and we use $t_1=0$. 
The outcome depends on the parameter of the 
dipole-dipole interaction $\chi$ and the amount of excess energy $\delta {\cal K}$. 
Here, we fix the dipole-dipole interaction and investigate the energy transfer as the energy given
to the first dipole increases. 
This dipole-dipole interaction parameter depends on the molecular species, through the
rotational constant and permanent 
electric dipole moment,  
and on the separation between the dipoles.
In this work, we use $\chi=10^{-5}$. In case we were considering the dipoles to be cold LiCs
molecules trapped in an optical lattice, the value $\chi=10^{-5}$ would
corresponds to an optical lattice with $a_l=429$~nm.
The parameter $\delta {\cal K}$ is given in units of $\chi$, \ie, 
in the energy units of the potential energy
surface $V_{\cal M}(\theta_1,\theta_2,t)$ for $\beta=0$, and 
we investigate the interval $2 \chi\le \delta {\cal K}\le  8 \chi$.
The final time is fixed to $t_f=5\times10^4$.
Our numerical tests have shown that
this value for the stopping time is appropriate for the correct
characterization of the outcomes.

\begin{figure}[t]
\includegraphics[scale=0.4]{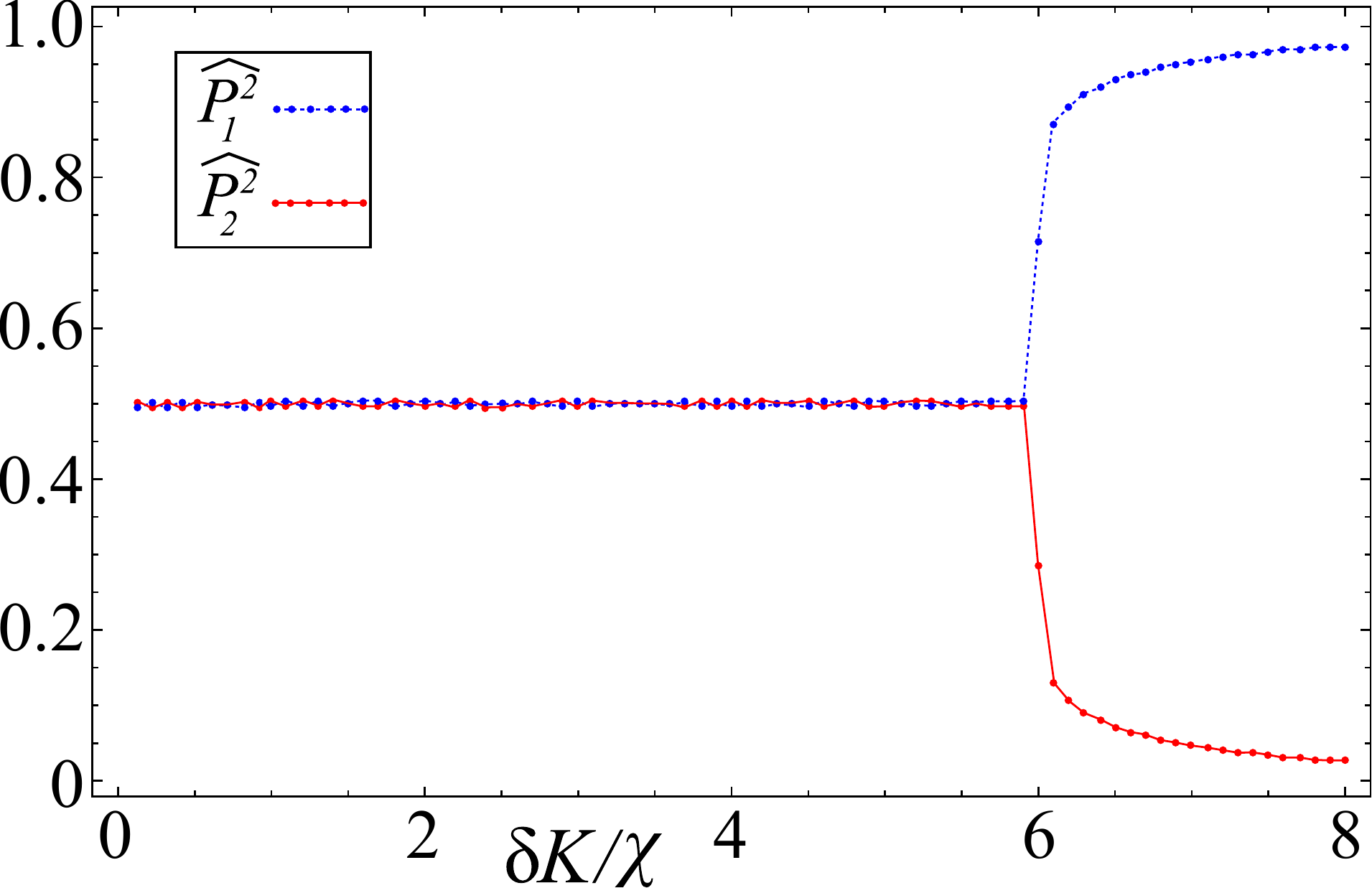}
\caption{For the field-free system, 
the normalized time-averaged kinetic energies of the dipoles
$\widehat{P_1^2}$ (blue thick solid line) and $\widehat{P_2^2}$ (red thin solid line), see~\autoref{average}, 
as a function of the initial excess energy of the first molecule  $\delta {\cal K}$. The dipole-dipole interaction
strength is $\chi=10^{-5}$. }
\label{fi:evolution}
\end{figure} 

\begin{figure}
\includegraphics[scale=0.4]{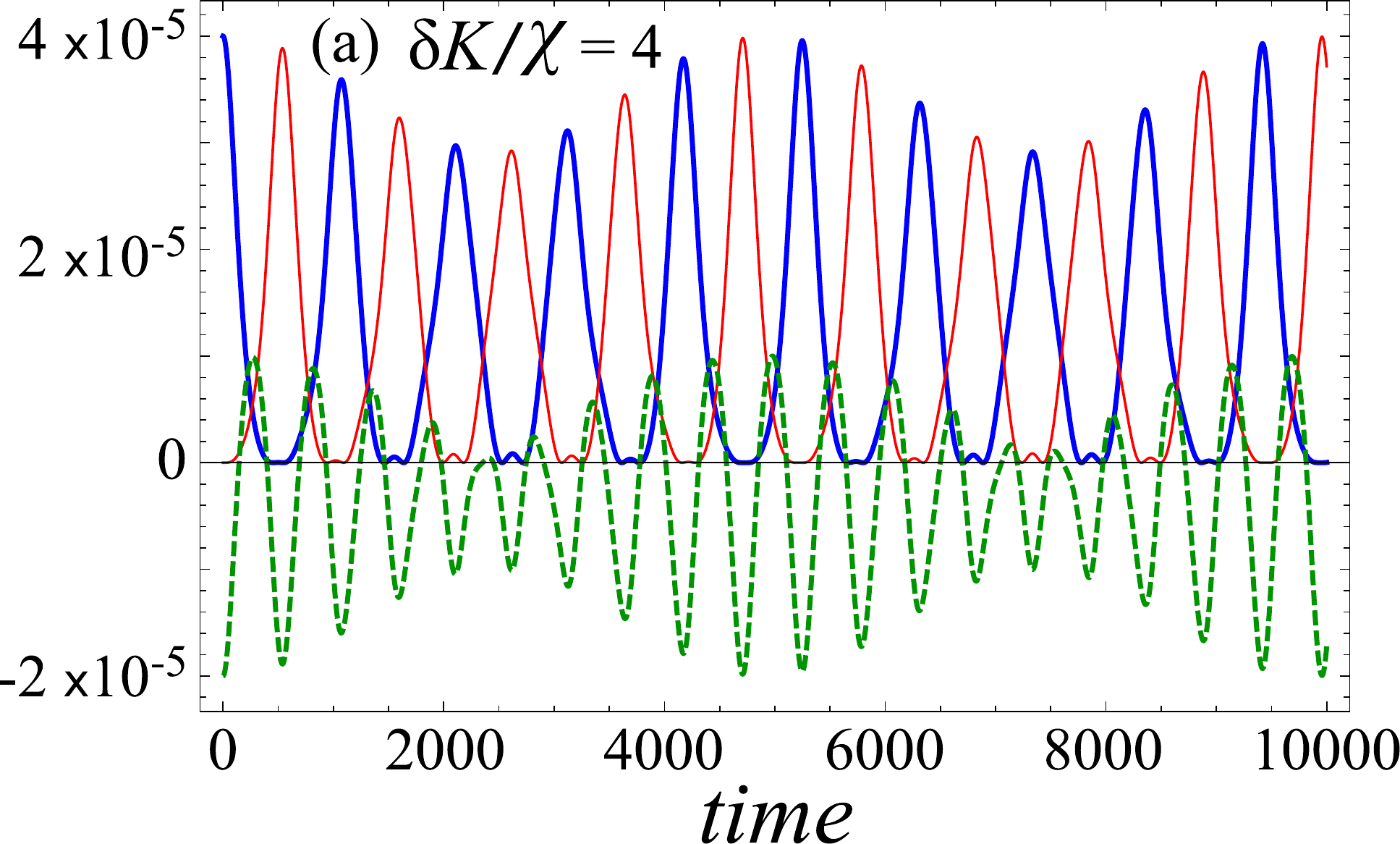}\\[3ex]
\includegraphics[scale=0.4]{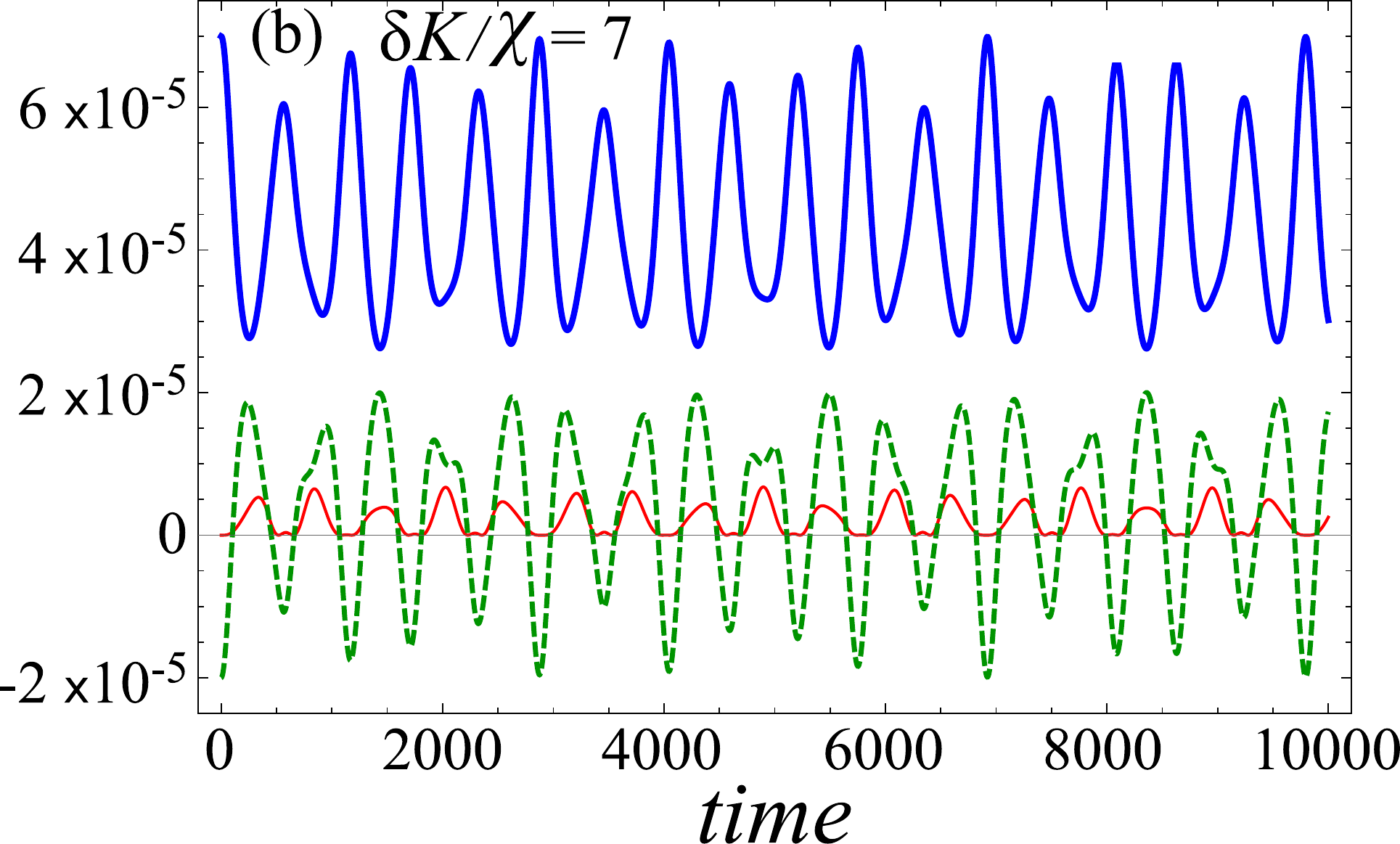}
\caption{For the field-free system, time-evolution of the kinetic
energies $P_1^2(t)$ (blue thick solid line) and $P_2^2(t)$ 
(red thin solid line)
and the potential energy $V_{\cal M}(\theta_1,\theta_2, t)$ (green dashed line).
The initial excess energies of the first molecule are (a) $\delta {\cal K}= 4 \chi$ and (b) $\delta {\cal K}= 7 \chi$.
The dipole-dipole interaction strength is $\chi=10^{-5}$.}
\label{fi:pxpy}
\end{figure} 
The   normalized time-averaged  kinetic energies of the dipoles 
are  shown in~\autoref{fi:evolution} as the excess energy $\delta {\cal K}$ increases. 
If the  excess energy $\delta {\cal K}$ is smaller than the critical value $\delta {\cal K}_c \approx 6\chi$,
the system is in an equipartition energy regime,  $\widehat{P_1^2}$ is very close to $\widehat{P_2^2}$,  
 and there is a continuous energy flow between the rotors. 
This behavior is illustrated  for $\delta {\cal K}= 4 \chi$ in~\autoref{fi:pxpy}a  with the
time evolution of the kinetic energies $P_1^2(t)$  and  $P_2^2(t)$,
and the potential energy $V_{\cal M}(\theta_1,\theta_2,t)$. 
 For  $\delta {\cal K}\approx 6\chi$, this equipartition regime
abruptly breaks and for  $\delta {\cal K}\gtrsim 6\chi$, most of the kinetic energy
remains in the first dipole.
As a consequence, the energy flow  between the dipoles is interrupted as shown in~\autoref{fi:pxpy}b
for $\delta {\cal K}= 7 \chi$.
The dynamics of this first rotor is essentially different in these two regimes.
For $\delta {\cal K}<6 \chi$, the kinetic energy $P_1^2(t)$ oscillates and  reaches zero as minimal value, 
see~\autoref{fi:pxpy}a, 
which indicates a non continuous rotation and this first dipole is  at rest  at these minima. 
For $\delta {\cal K}>6 \chi$, 
$P_1^2(t)>0$, cf.~\autoref{fi:pxpy}b,  which indicates that  the first dipole is performing a continuous rotation. 
In contrast, the smaller kinetic energy of the second rotor $P_2^2(t)$ oscillates
with a non continuous rotation  and  has as minimum value zero in both regimes.
This behavior in the energy flux between the dipoles was already observed by de Jonge {\sl et al.}
\cite{Ratner3}. In that paper, the authors provide an analytical explanation showing that the energy
transfer is only possible in a low energy regime.

\begin{figure*}
\includegraphics[scale=0.64]{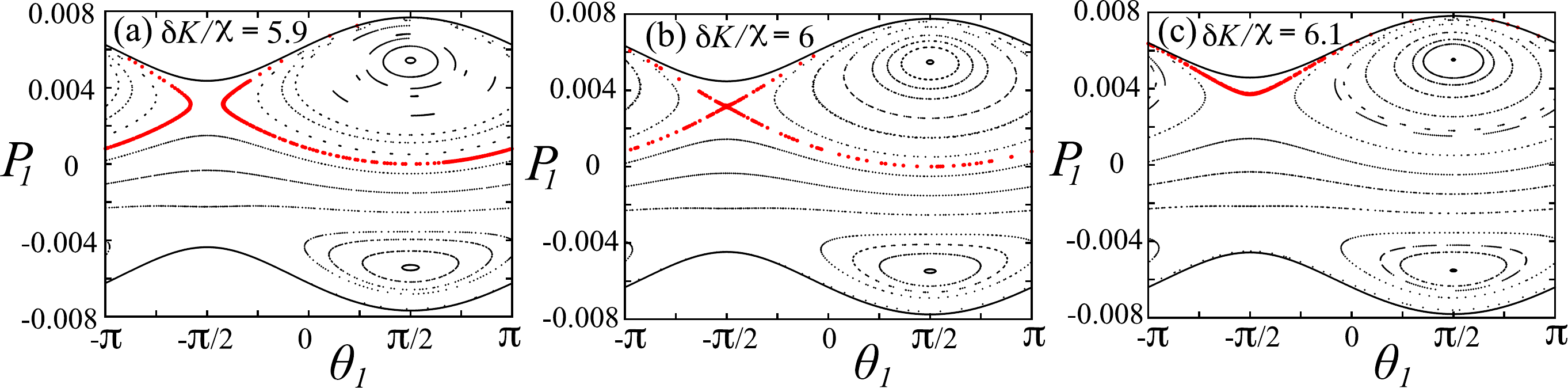}
\caption{For the field-free system with dipole-dipole interaction $\chi=10^{-5}$, Poincar\'e surfaces of section in the plane
$(P_1, \theta_1)$ with $\theta_2=\pi/2$ for three different initial excess energies  in the
neighborhood of the critical value $\delta {\cal K}_c=6 \chi$. The red  thick points correspond to the trajectory 
$\tau$  with initial conditions~\eqref{ci}. }
\label{fi:sos3}
\end{figure*} 
To gain a deeper physical insight into the energy transfer mechanism,
we present in~\autoref{fi:sos3}
 the Poincar\'e surface of section in the plane $(P_1, \theta_1)$ with $P_2=0$ for three initial excess energies. 
This  surface of section provides a good illustration for the trajectory $\tau$  with initial conditions~\eqref{ci}. 
For $\delta {\cal K}_c<6 \chi$, the orbit $\tau$ has a vibrational nature, see~\autoref{fi:sos3}a, whereas 
we observe in~\autoref{fi:sos3}c that  its nature  becomes rotational for $\delta {\cal K}_c>6 \chi$.
At the critical value $\delta {\cal K}_c=6 \chi$, the $\tau$ orbit is the separatrix, cf.~\autoref{fi:sos3}b, 
that keeps  rotational and vibrational regions away from each other.
That is why  the transition from the energy equipartition regime to the non-equipartition occurs at the critical 
value $\delta {\cal K}_c=6 \chi$.

These energy transfer mechanisms can be explained analyzing the dynamics of the two 
pendula in the rotated reference frame.
The momenta  in the LFF and in the rotated frame are related according to
 \begin{equation}
P_1 = \frac{P'_1- P'_2}{\sqrt{2}},   \qquad   P_2 =\frac{P'_1+P'_2}{\sqrt{2}},
\end{equation}
and the time-averaged kinetic energy of each dipole can be written as
 \begin{equation}
 \label{avekinetic}
\langle P_1^2 \rangle = \frac{\langle P'^2_1\rangle +\langle P'^2_2 \rangle}{2} -  \langle P'_1 P'_2 \rangle,
\end{equation}
 \begin{equation}
  \langle P_2^2 \rangle = \frac{\langle P'^2_1\rangle +\langle P'^2_2 \rangle}{2} +  \langle P'_1 P'_2 \rangle,
\end{equation}
\noindent
that is, the  time-averaged kinetic energies of the dipoles  differ by  twice the time-average of the
product  of momenta of the pendula $\langle P'_1 P'_2 \rangle$.

In the rotated reference frame, using the transformations~\eqref{rotation}, the
initial conditions $(\theta_1(0), P_1(0), \theta_2(0), P_2(0))$ give rise 
to two (uncoupled) pendular motions governed by the Hamiltonians~\eqref{rotation2} with
energies $E_1'$ and $E_2'$. These energies determine the  motion in the rotated frame, and
the kinetic energy transfer mechanism between the rotors.
If the energies of the pendula are smaller than the maxima of the potentials $V'_{1}(\theta_1')$ and
$V'_{2}(\theta_2')$, \ie,  $E_1'<3\chi/2$ and $E_2'<\chi/2$, respectively,  the total energy of the
system is $E=E_1'+E_2'<2 \chi$, and  both
pendula describe periodic oscillations, \ie, the 
momenta $P'_1$ and $ P'_2$ are periodic functions around zero with 
 $\langle P'_1 \rangle=\langle P'_2 \rangle=0$,  and the time-averaged product
$\langle P'_1 P'_2 \rangle$ is zero. As a consequence, 
$\langle P_1^2 \rangle=\langle P_2^2 \rangle$, which means that 
the system will always belong to the equipartition kinetic energy regime.
The same behavior occurs  when $E_1'>3\chi/2$ or $E_2'>\chi/2$,  and
at least one of them
is in the vibrational regime with $\langle P'_i\rangle=0$, whereas the other one
describes complete periodic rotations and its 
momentum is a periodic function around a nonzero value having a nonzero time average,  and again 
it holds $\langle P'_1 P'_2 \rangle=0$.
Finally,  if the initial condition leads to a pendular energy distribution with $E_1'>3\chi/2$ and $E_2'>\chi/2$, then, 
both pendula are in the rotational regime,  the time-average product
$\langle P'_1 P'_2 \rangle$ is nonzero, and the equipartition regime is not met.

\begin{figure*}
\includegraphics[scale=0.27]{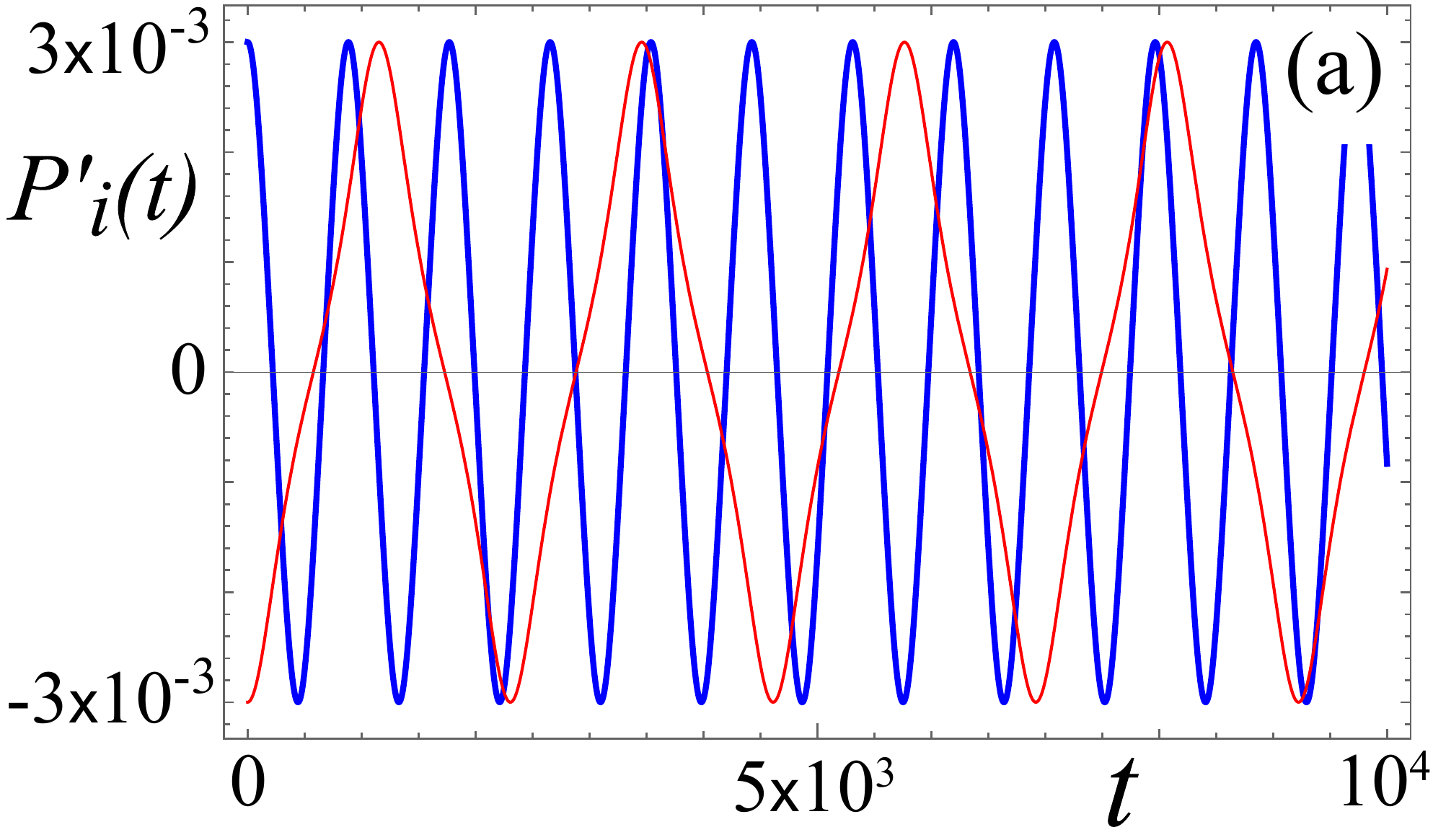} \quad
\includegraphics[scale=0.27]{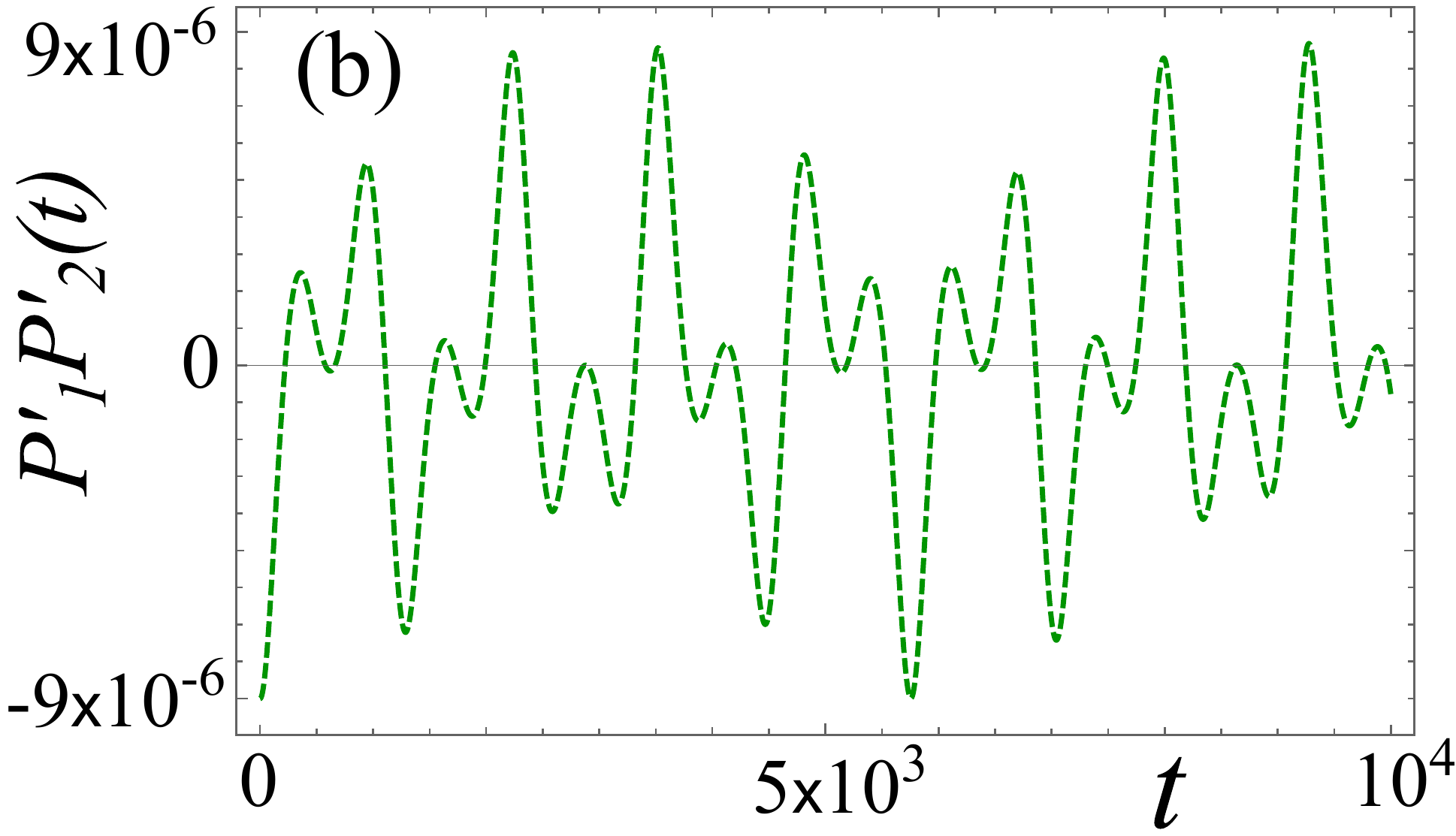} \\[2ex]

\includegraphics[scale=0.27]{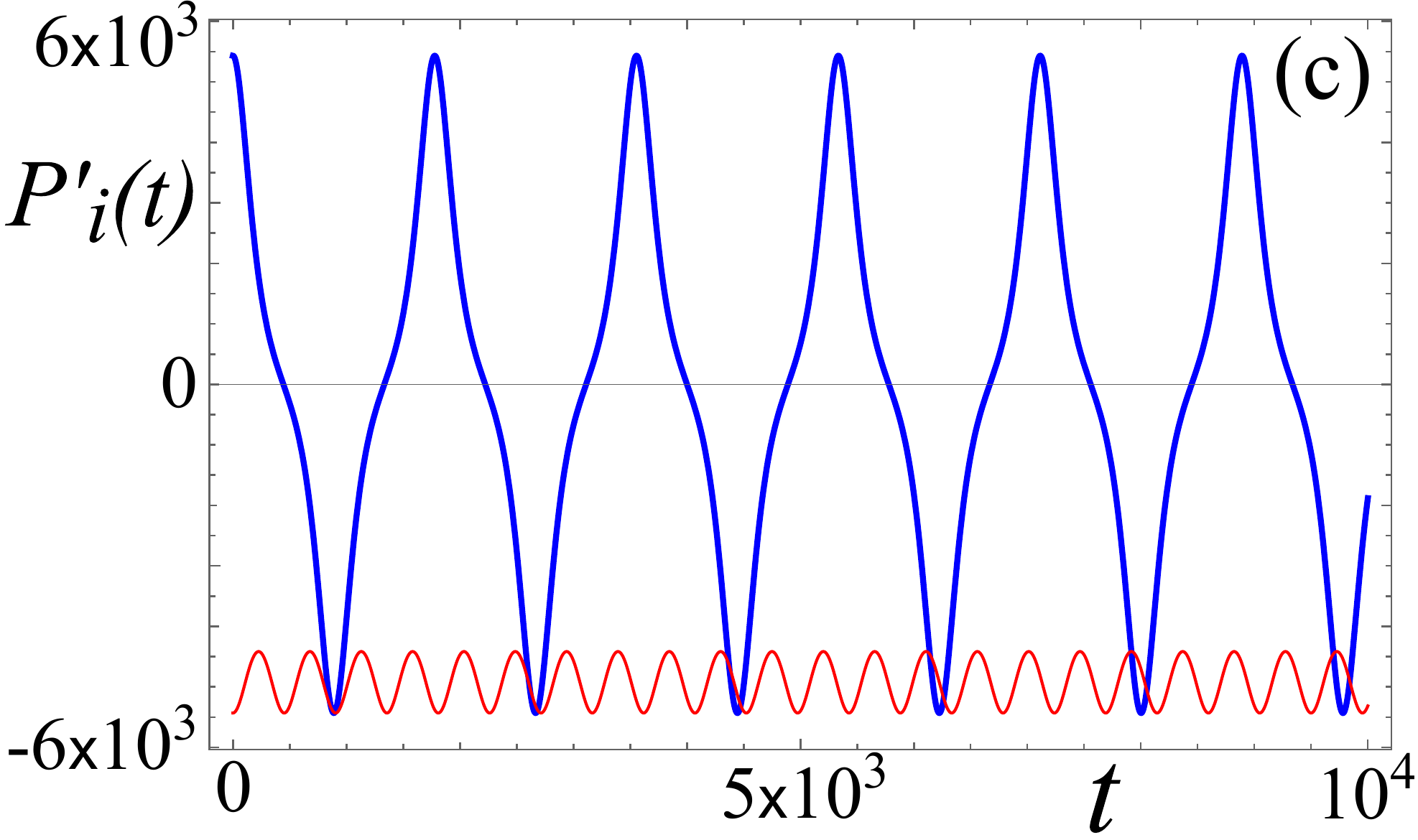} \quad
\includegraphics[scale=0.27]{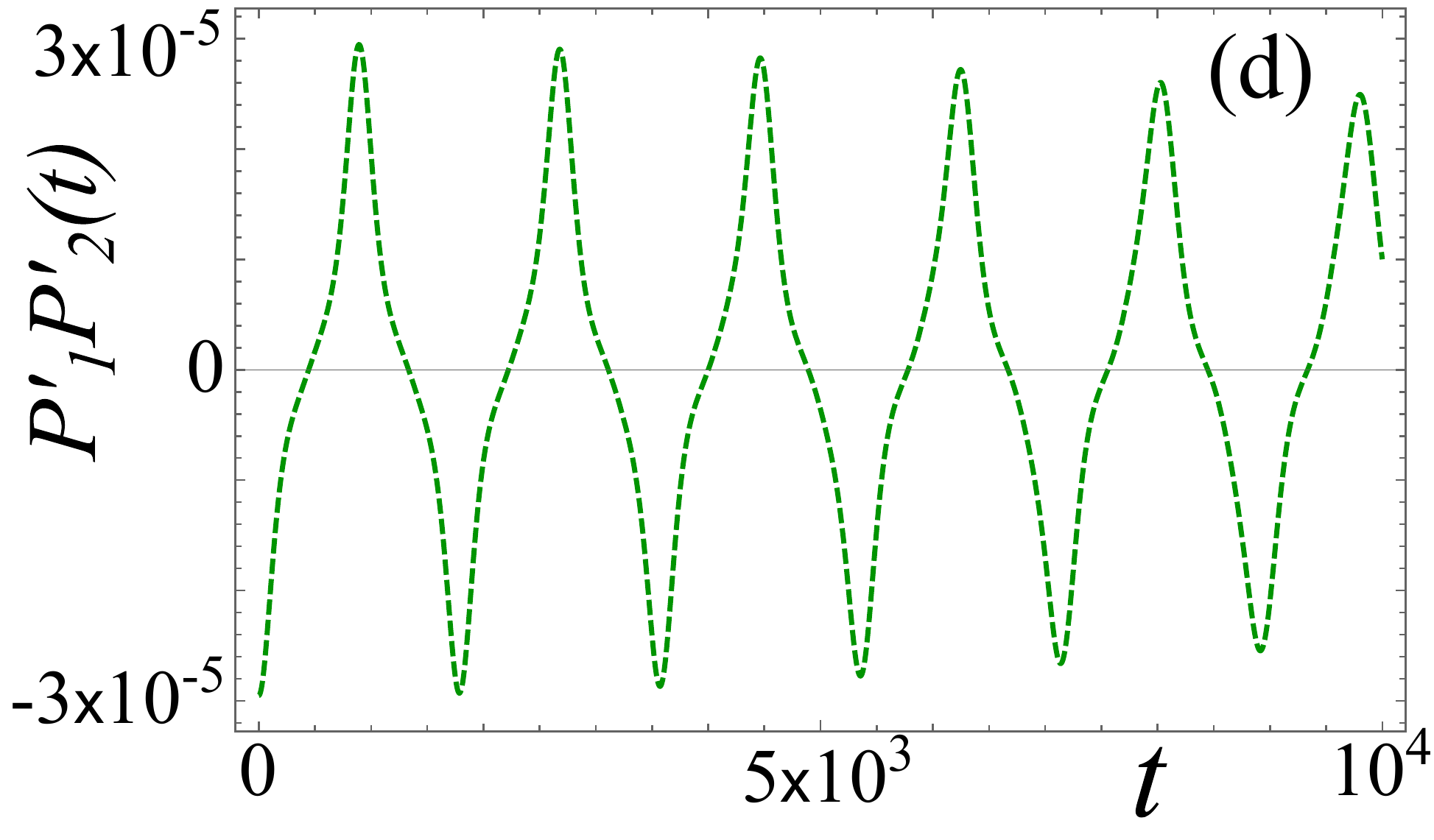}\\[2ex]

\includegraphics[scale=0.27]{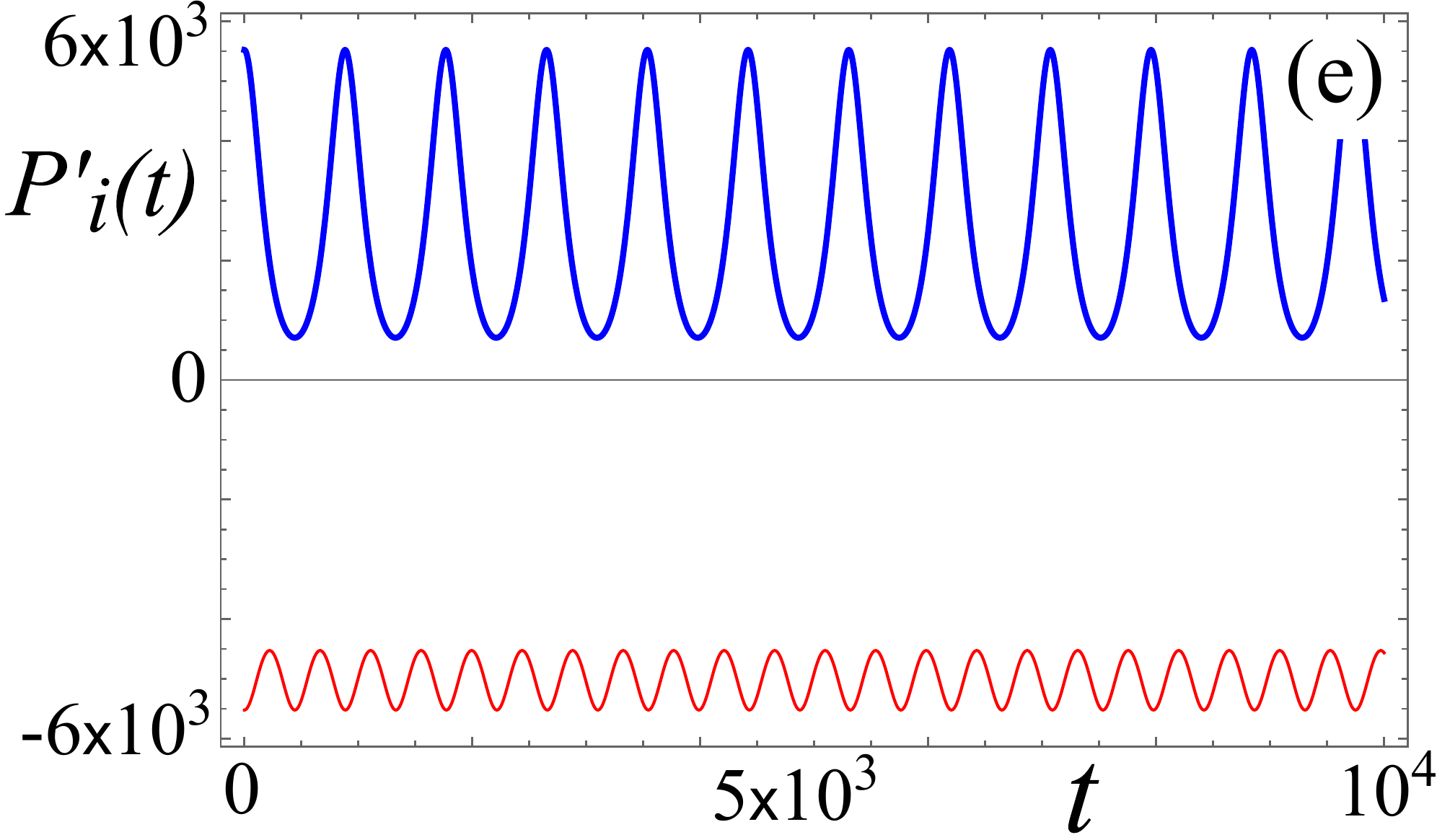}  \quad
\includegraphics[scale=0.27]{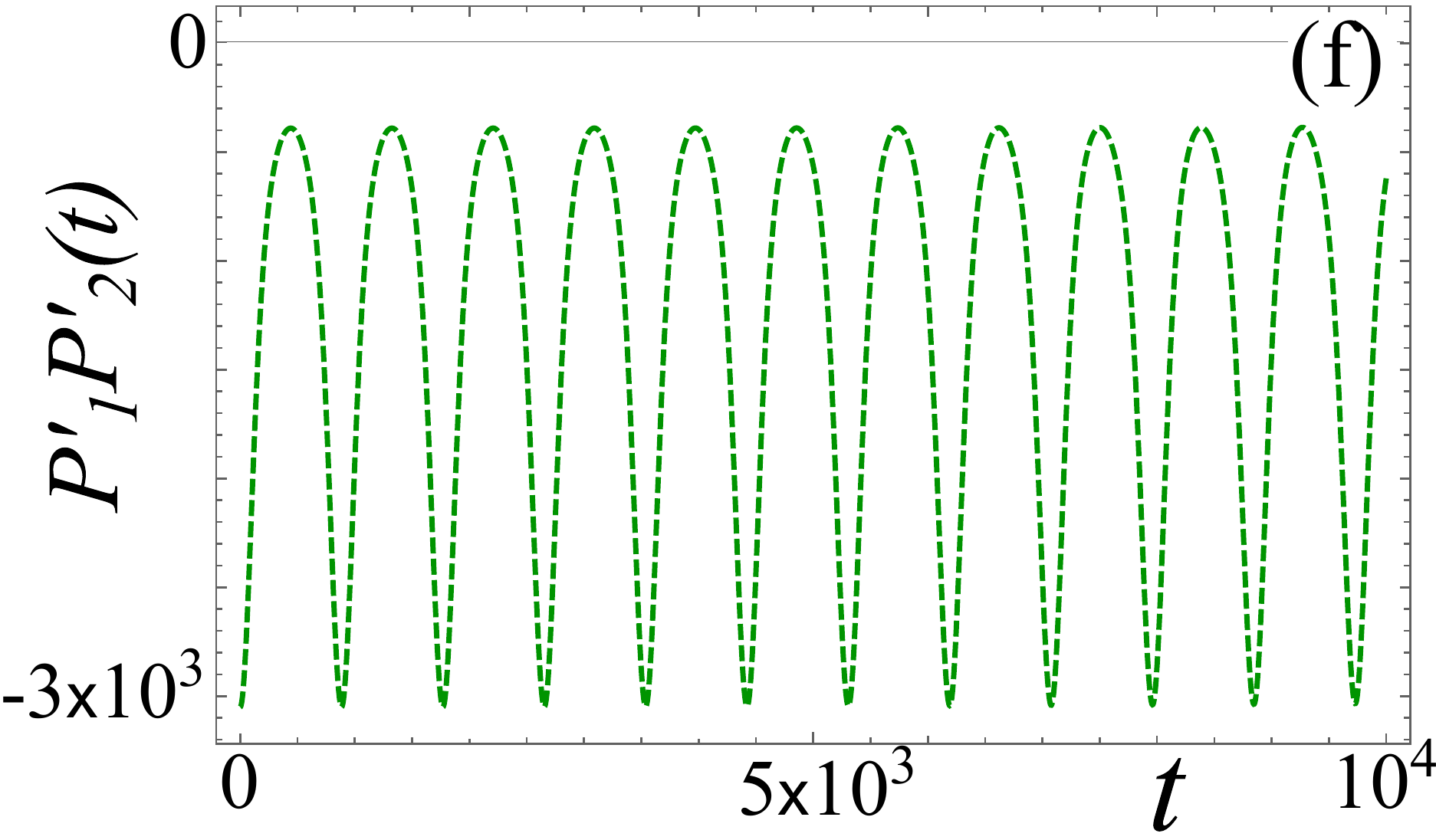} 
\caption{For the field-free system, time evolution of the momenta $P_1'^2(t)$ (blue thick solid line) and $P_2'^2(t)$ 
(red thin solid line)
of the uncoupled pendula (upper row) and the product $P'_1(t)P'_2(t)$ (green dashed line) (lower row) for the excess energies (a) and (b) $\delta {\cal K}=1.8\chi$; 
(c) and (d) $\delta {\cal K}=5.9\chi$; (e) and (f) $\delta {\cal K}=6.1\chi$, for $P'_1(t)$ and $P'_2$(t), and
$P'_1(t)P'_2(t)$, respectively. The dipole-dipole 
interaction strength is  $\chi=10^{-5}$.}
\label{fi:timevolP}
\end{figure*}

For the orbit  $\tau$, the initial conditions~\eqref{ci} expressed  in the rotated frame read
\[
\theta'_1(0)=\frac{\pi}{\sqrt{2}},\,  \theta'_2(0)=0,\,
P'_1(0)=\sqrt{\frac{\delta {\cal K}}{2}},  \, P'_2(0)=-\sqrt{\frac{\delta {\cal K}}{2}}
\]
In this rotated frame,  the excess kinetic energies in the pendula
are the same, $\delta {\cal K}/2$, whereas their energies are 
\begin{equation}
\label{hamilpen}
E_1'=\frac{\delta {\cal K}}{2}-\frac{3}{2} \chi, \qquad E_2'=\frac{\delta {\cal K}}{2}-\frac{1}{2} \chi.
\end{equation}
\noindent
If the excess energy satisfies 
$\delta {\cal K}<2\chi$, both pendula are in an oscillatory motion, and the dipoles belong to the
energy equipartition regime. This is illustrated  for $\delta{\cal K}=1.8 \chi$ 
in~\autoref{fi:timevolP}a and~\autoref{fi:timevolP}b with  the
time evolution of $P_1'(t)$, $P_2'(t)$ and  $P_1' (t)P_2'(t)$, respectively.
If the excess energy satisfies 
$2\chi<\delta {\cal K}<6\chi$, the second pendulum performs complete rotations, whereas
the first one still performs a vibrational motion, see in~\autoref{fi:timevolP}c and~\autoref{fi:timevolP}d
the evolution of the momenta and the product of momenta for  $\delta{\cal K}=5.9 \chi$.
In this situation, the dipole  relaxes again to the equipartition regime. However, if the
excess energy is $\delta {\cal K}>6\chi$, both pendula have a rotational motion and the dipoles do not reach 
the equipartition regime, as an example see for $\delta{\cal K}=6.1 \chi$
the time-evolution of the momenta and the product of momenta in~\autoref{fi:timevolP}e and~\autoref{fi:timevolP}f.

\section{Energy Transfer in an External Electric Field}
\label{sec:en_tra_dc_field}

In this section we explore the energy transfer between the two dipoles in the presence of an external electric field.
Again, we assume that the dipoles are initially in the stable head-tail configuration with zero kinetic energy. 
At time $t=0$,  a certain
amount  of kinetic energy $\delta {\cal K}$ is given to the first dipole, and simultaneously the 
electric field is turned on with  the linear profile~\eqref{pulse}. 
Using the initial conditions~\ref{ci}, the equations of motion~\ref{ecumoviA} are integrated up
to a final time $t_f$, and
we  compute the normalized time-averaged momenta $\widehat P_1^2$ 
and $\widehat P_2^2$ from~\autoref{average}. 
As in the field-free system, we are using  a dipole-dipole interaction with strength $\chi=10^{-5}$, and a
final time 
$t_f=5\times10^4$.
The strength of the electric field is varied in the interval $0.01\chi\le \beta\le1000\chi$.
We assume a switched-on  time for the field of $t_1=1200$, 
which roughly corresponds to  $100~ns$,
that could be achieved in current experiments with realistic field strengths.

\begin{figure}[b]
\includegraphics[scale=0.41]{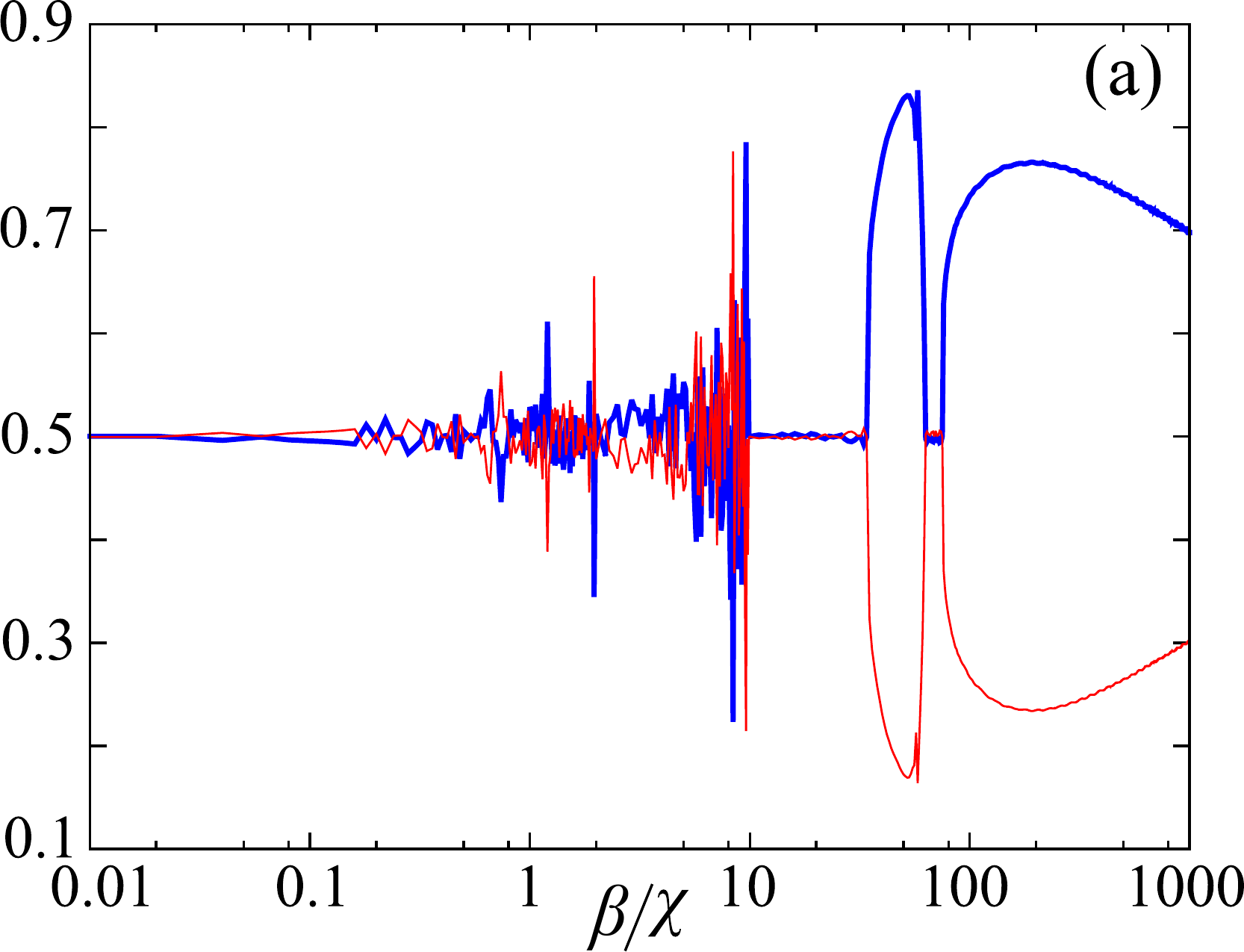} 
 \includegraphics[scale=0.41]{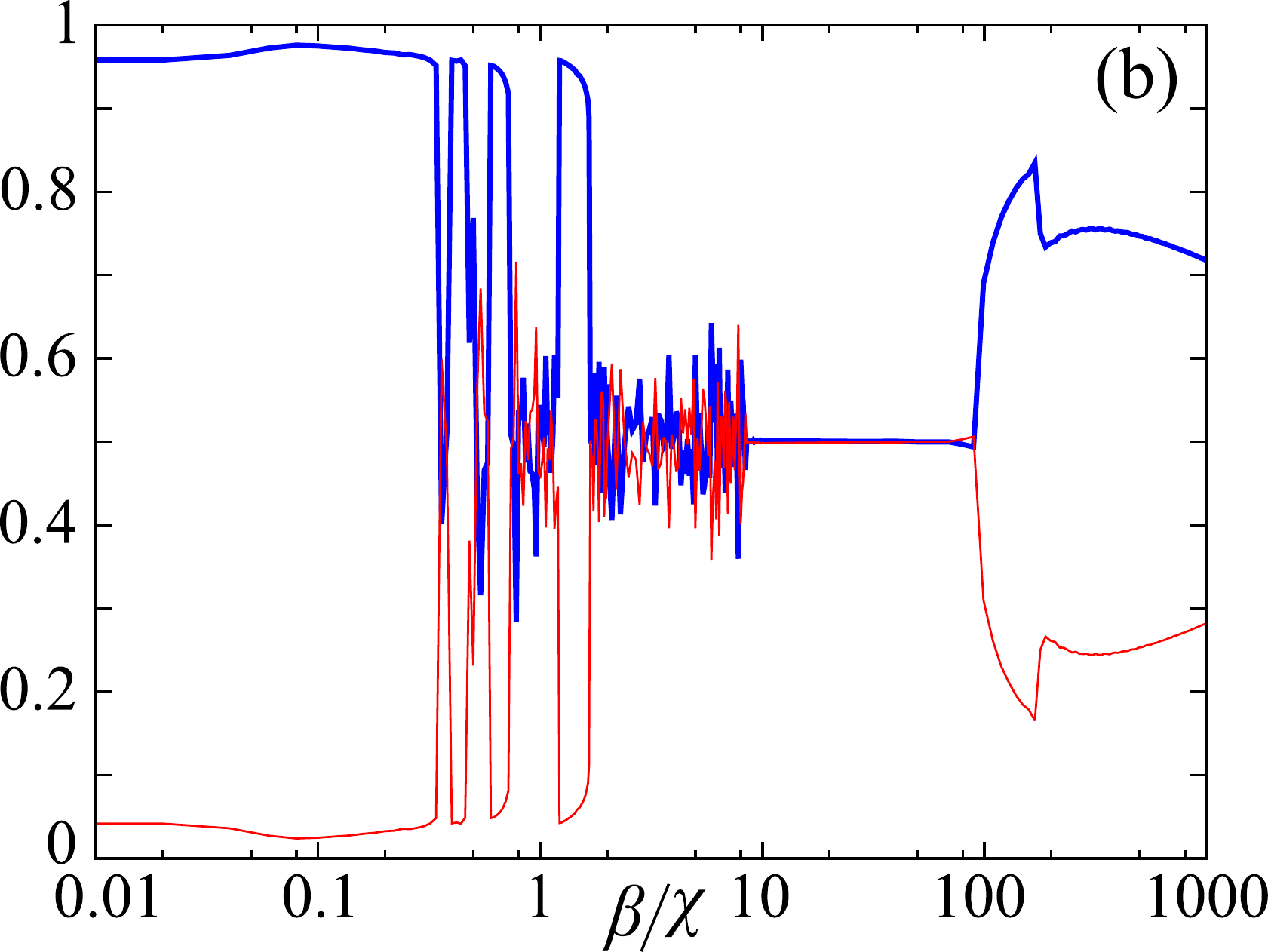} 
\caption{The normalized time-averaged kinetic energies of the dipoles
$\widehat{P_1}$ (blue thick solid line) and $\widehat{P_2}$ (red thin solid line)
as a function of the ratio between the  electric field parameter $\beta$ and the dipole-dipole parameter $\chi$ 
for two initial excess energies of
the first rotor (a) $\delta {\cal K}=4\chi$ and (b) $\delta {\cal K}=7\chi$. The dipole-dipole 
interaction parameter is  $\chi=10^{-5}$.}
\label{fi:evolution2}
\end{figure}
Based on the results for the field-free system, we investigate the time-averaged kinetic energies 
$\widehat{P_1^2}$ and $\widehat{P_2^2}$ for $\delta {\cal K}=4\chi$
and $\delta {\cal K}=7\chi$ as  the field parameter varies. The results are depicted in~\autoref{fi:evolution2}. 
For these two excess energies,   $\widehat{P_1^2}$ and $\widehat{P_2^2}$ follow essentially different
behaviors as $\beta$ increases, but four common patterns can be identified in the two cases
 of~\autoref{fi:evolution2}.
 For small values $\beta\lesssim0.5\chi$, the dipole-dipole interaction is dominant and adding the external electric field has no relevant effect. As a consequence, we encounter the energy partition regimes for
  $\delta {\cal K}=4\chi$  (equipartition for $\beta=0$) and $\delta {\cal K}=7\chi$ (non-equipartition  for $\beta=0$). 
By increasing  the electric field in the interval  $0.5\chi\lesssim\beta\lesssim10\chi$, the energy partition 
 diagrams show sudden (random) variations. In this field range, 
 the dipole-dipole and electric field interactions are comparable in magnitude and  the system dynamics
 is sensitive to the variations of the the electric field parameter.
 For intermediate strengths,  the dipoles relax to an energy equipartition regime:
see  $10\chi \lesssim\beta\lesssim 40\chi$ and $10\chi\lesssim\beta\lesssim100\chi$,
for $\delta {\cal K}=4\chi$
and $\delta {\cal K}=7\chi$, respectively.
 Finally, for stronger electric fields, the system falls out of the equipartition regime, and  most
 of the kinetic energy remains in 
 one of the dipoles. For the initial conditions investigated here, most of the kinetic
 energy remains in the first dipole. By varying 
 the initial conditions, the role played by the two rotors could change, and the second rotor
 could store most of the kinetic energy.

\begin{figure}
\includegraphics[scale=0.38]{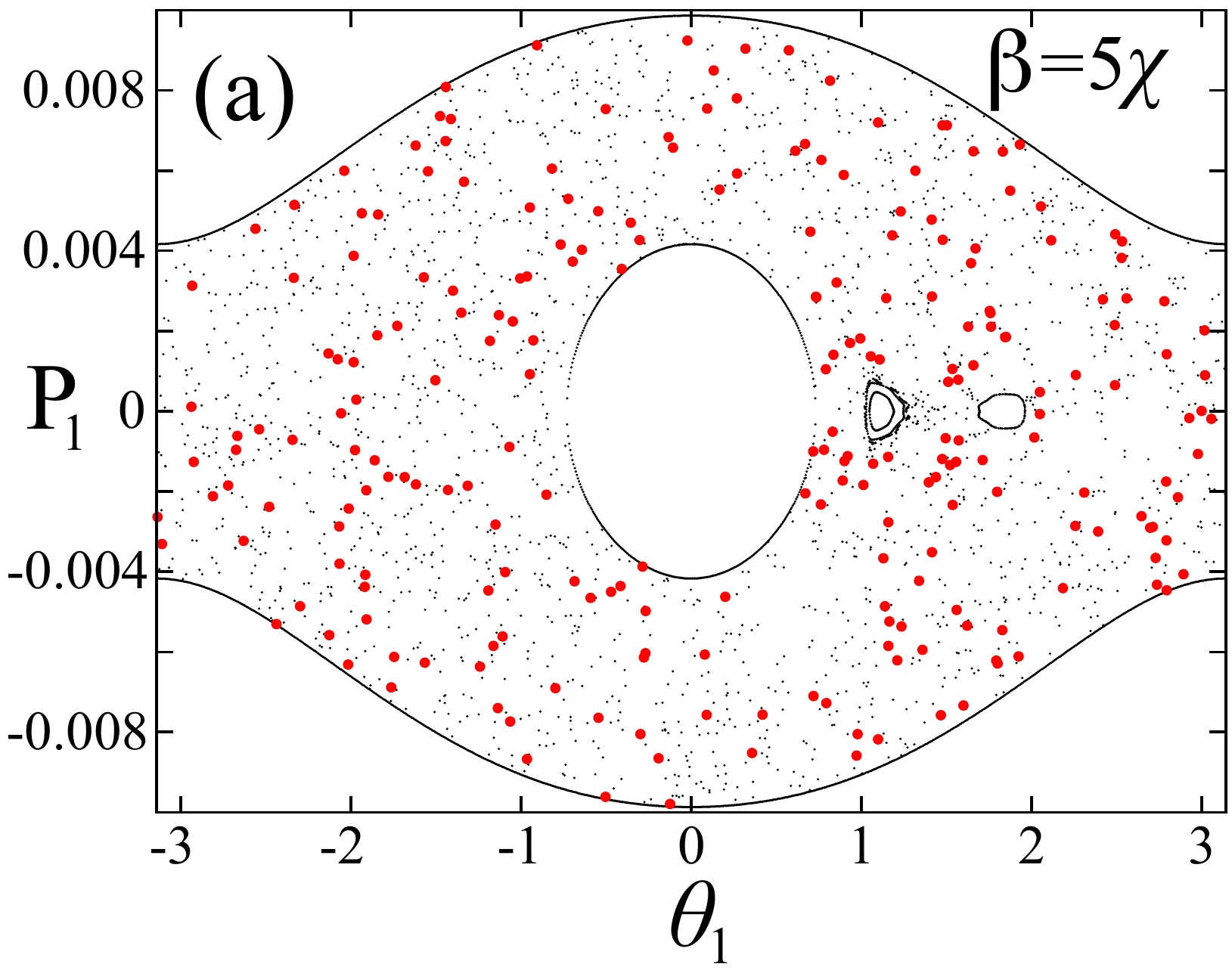} 
\includegraphics[scale=0.38]{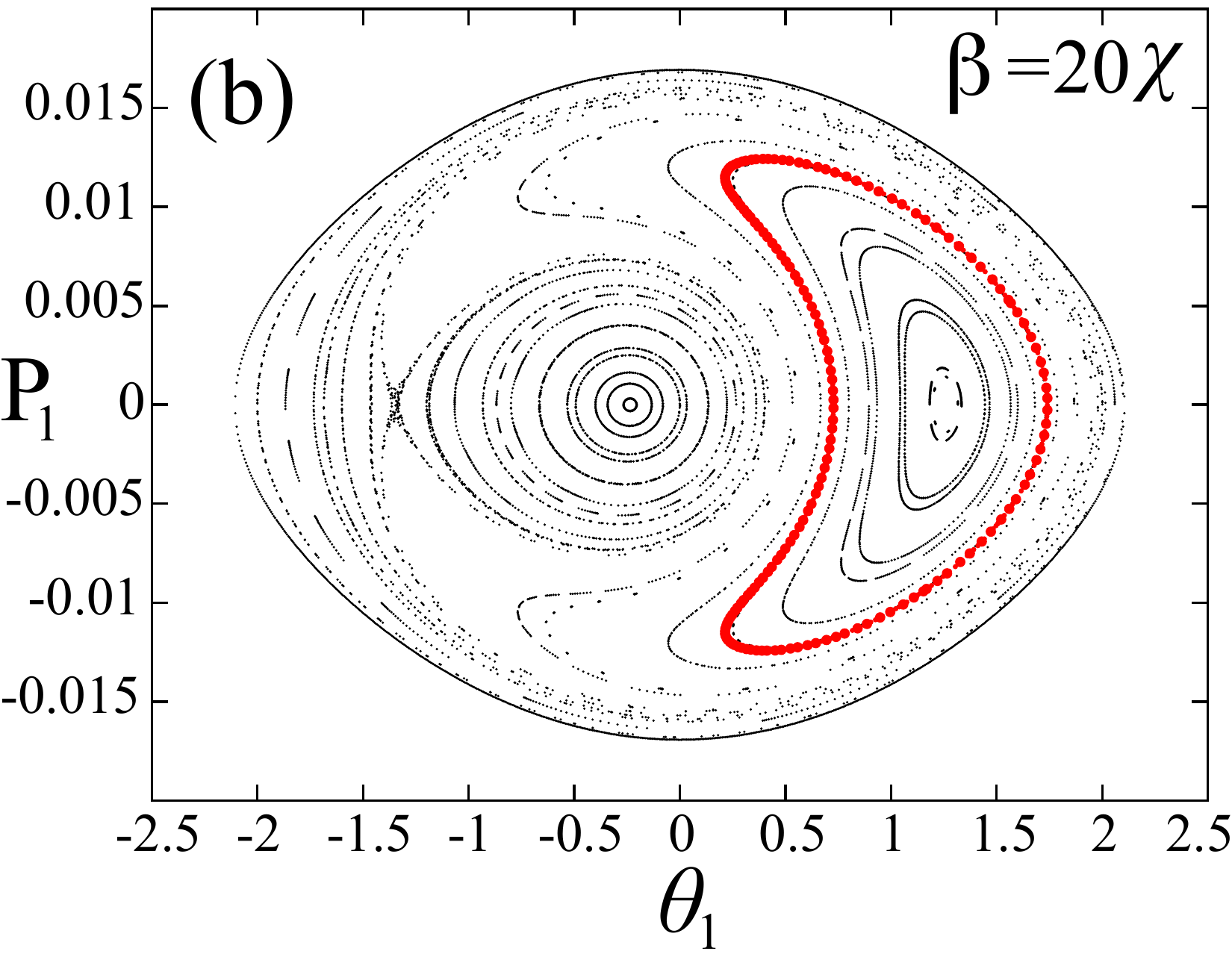} 
\includegraphics[scale=0.38]{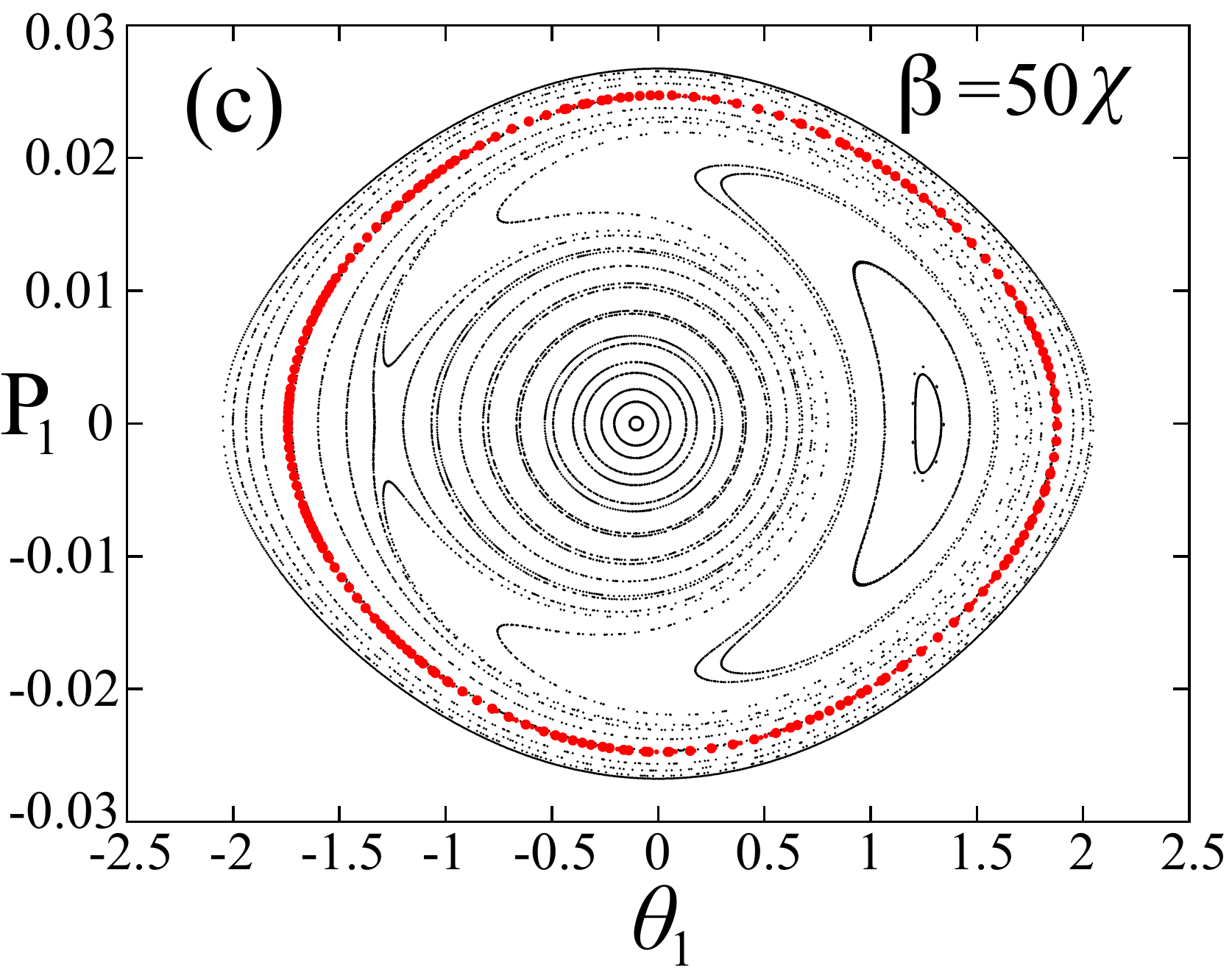} 
\includegraphics[scale=0.38]{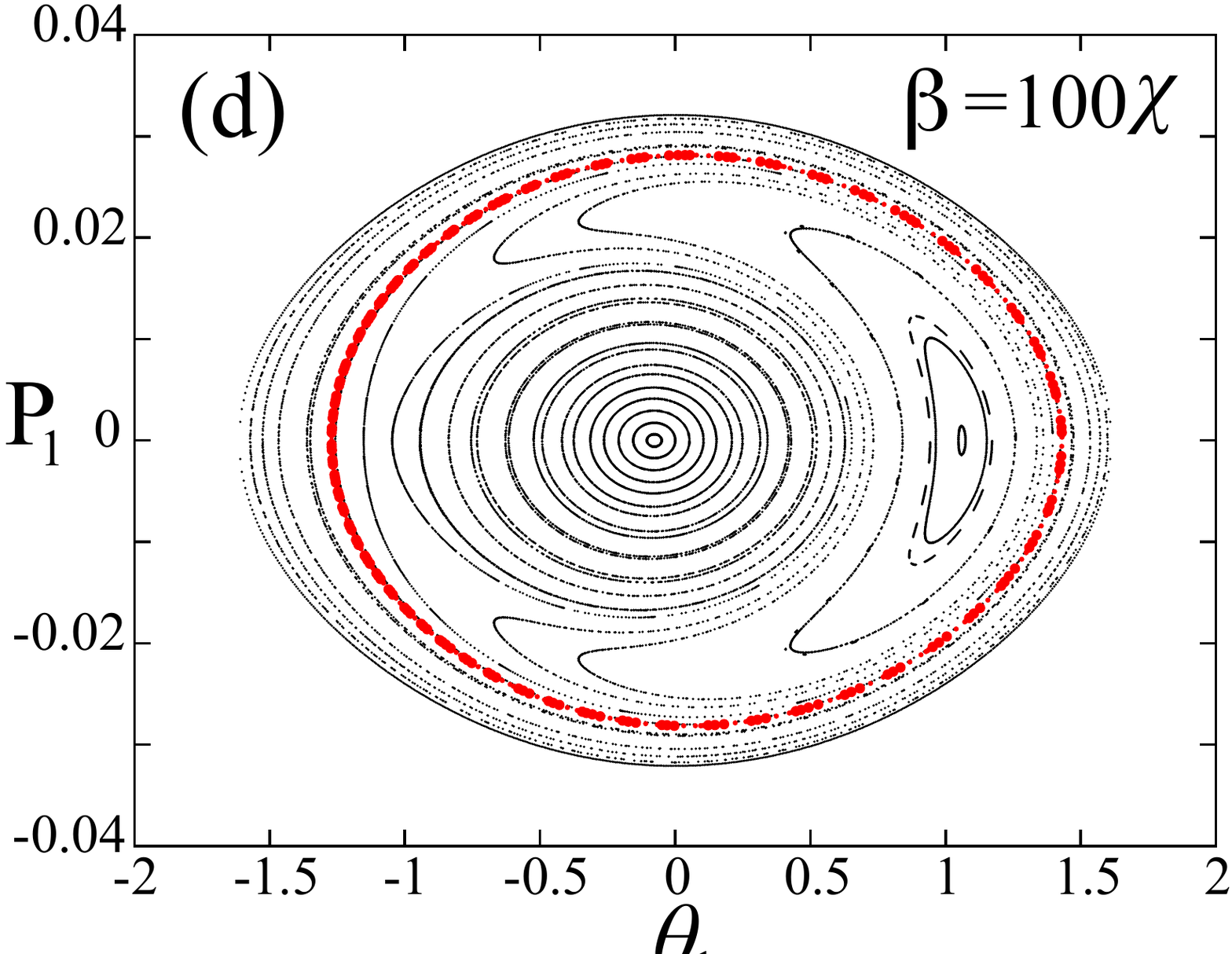}
\caption{Poincar\'e surface of
 section in the plane $(P_1, \theta_1)$ with $P_2=0$ for different values of the
 electric field parameter $\beta$. The dipole-dipole interaction
is  $\chi=10^{-5}$ and the  excess kinetic energy of the first dipole  is $\delta {\cal K}=4\chi$.
The thick (red) points correspond to the trajectory 
$\tau$ with initial conditions~\eqref{ci}.}
\label{fi:sos4}
\end{figure}
\begin{figure}
\includegraphics[scale=0.38]{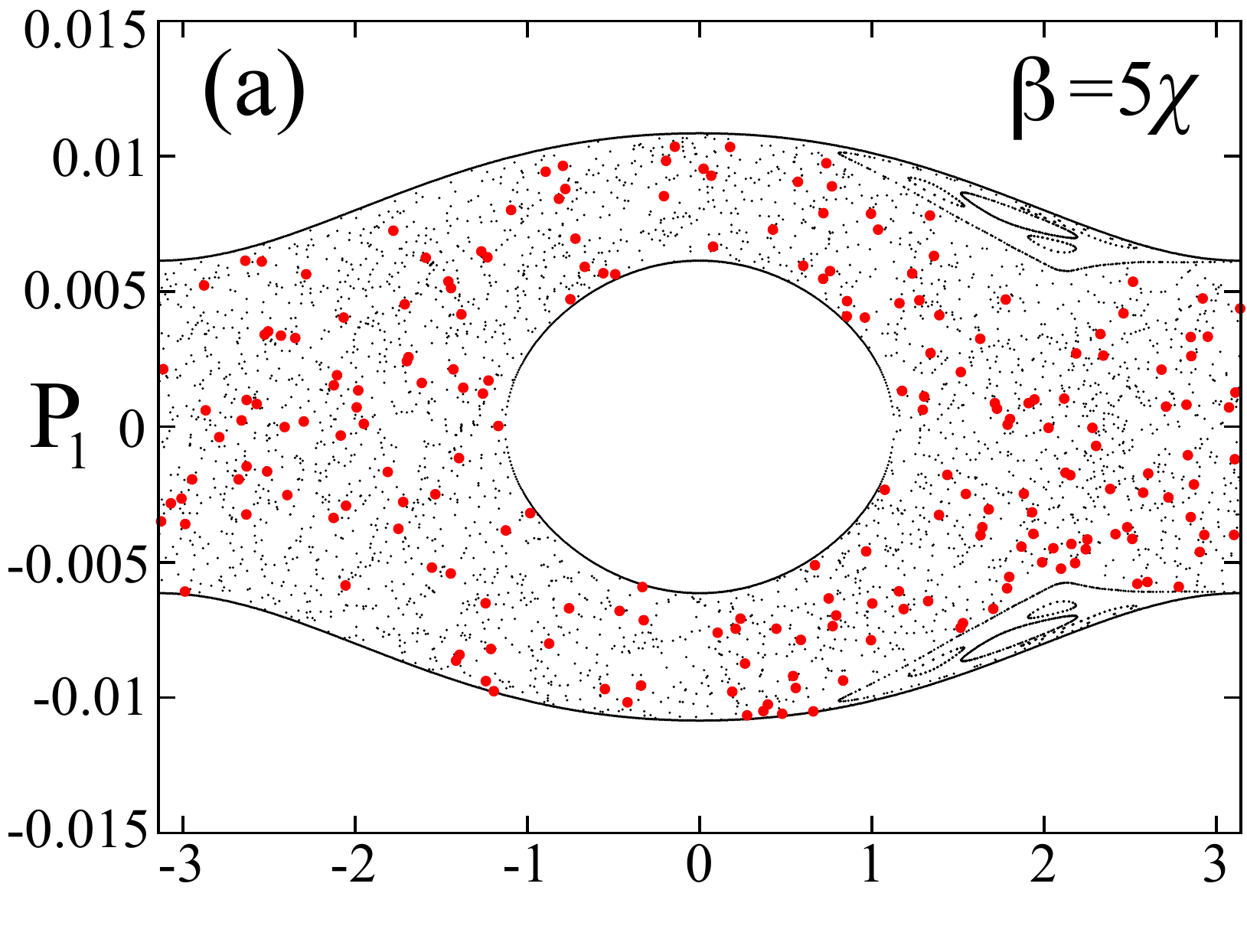}  \includegraphics[scale=0.38]{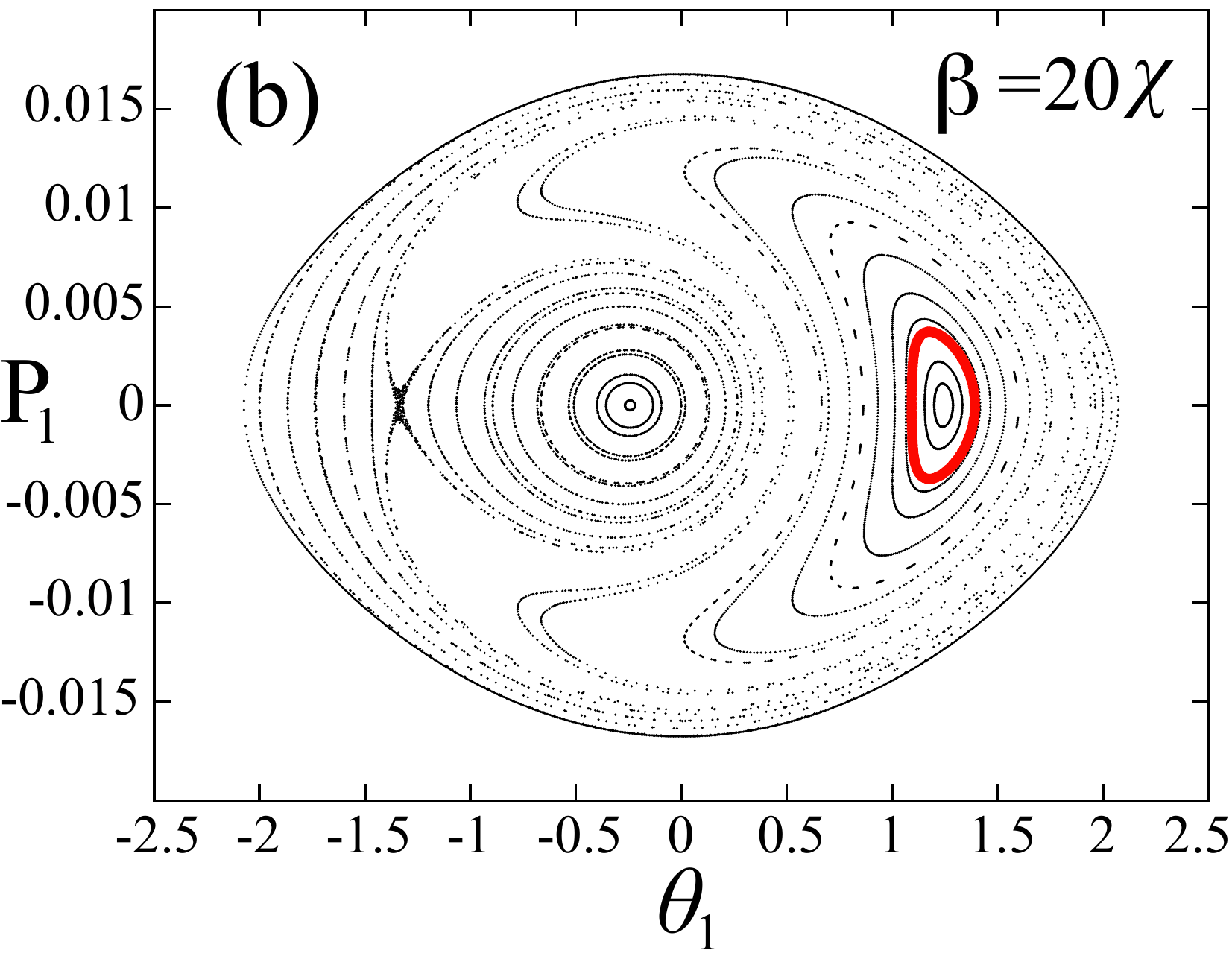}
\includegraphics[scale=0.38]{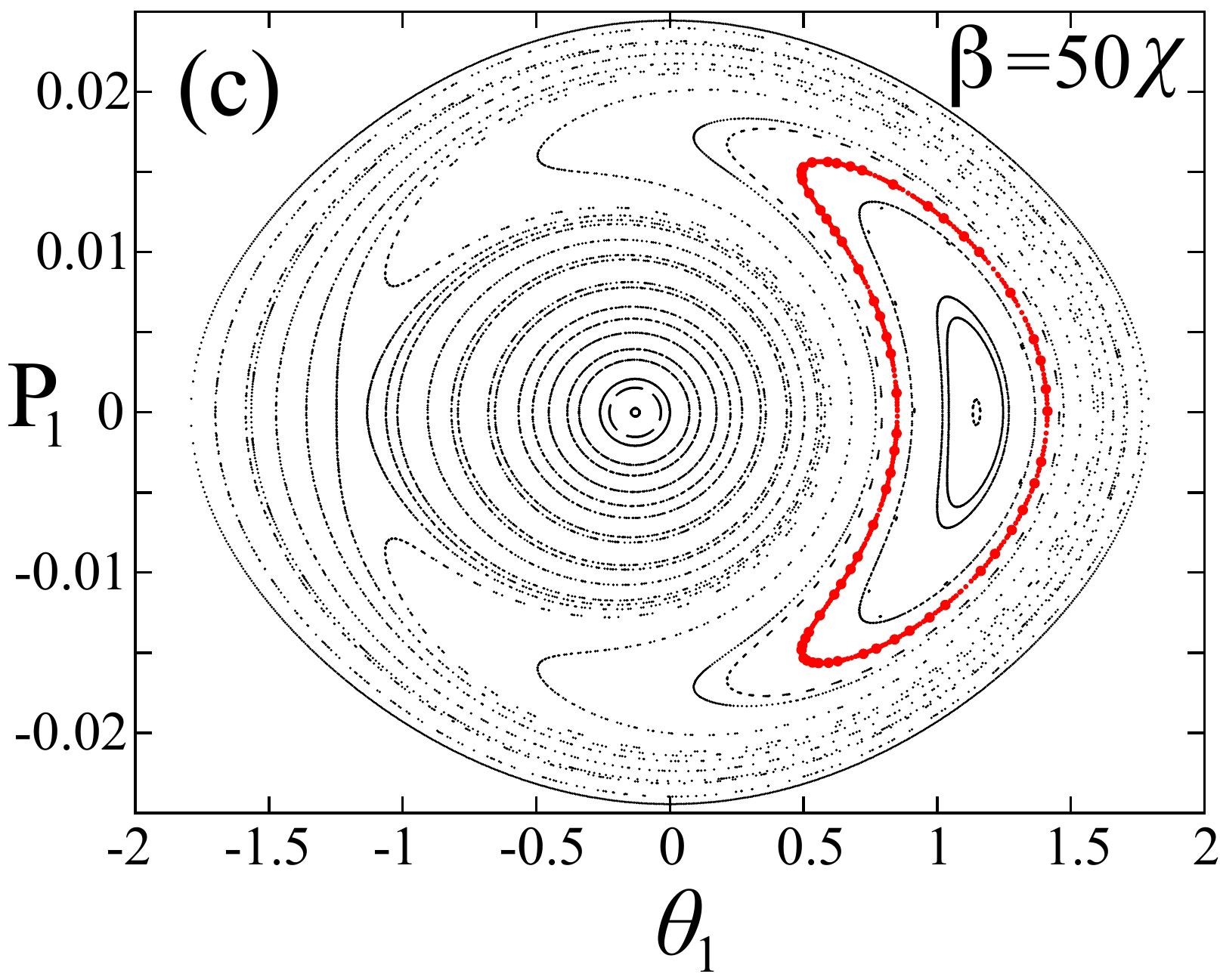} \includegraphics[scale=0.38]{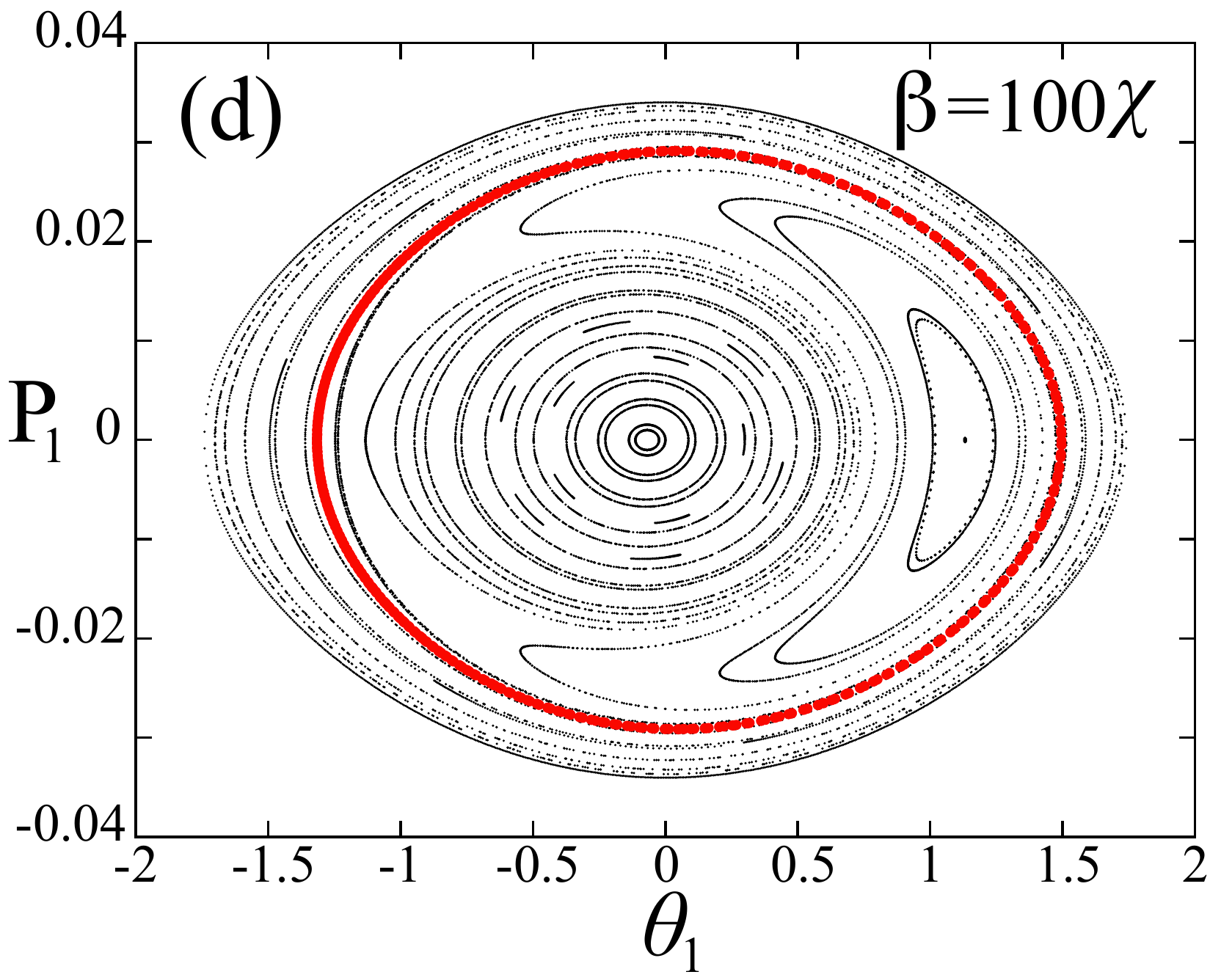} 
\caption{
The same as in~\autoref{fi:sos4}, but for an excess energy of the first dipole of  $\delta {\cal K}=7\chi$.}
\label{fi:sos7}
\end{figure}

The Poincar\'e surfaces of section provide a global picture of  the phase space
structure and are therefore suited to analyze and understand the kinetic energy transfer.
We analyze the Poincar\'e surfaces of section for a fixed $t>t_1$,  once the  electric field parameter has reached
its maximal strength $\beta$,  and the energy is constant.
To illustrate the trajectory $\tau$ with initial conditions~\eqref{ci}, a suitable surface of section for the Poincar\'e map is 
given by the intersection of the phase space trajectories with the plane
$(P_1, \theta_1)$ with $P_2=0$. In~\autoref{fi:sos4} and~\autoref{fi:sos7},
we show these Poincar\'e surfaces of  section for different values of the electric field parameter $\beta$ and for
the excess energies $\delta {\cal K}=4\chi$ and
$\delta {\cal K}=7\chi$, respectively.

\begin{figure*}
\includegraphics[scale=0.36]{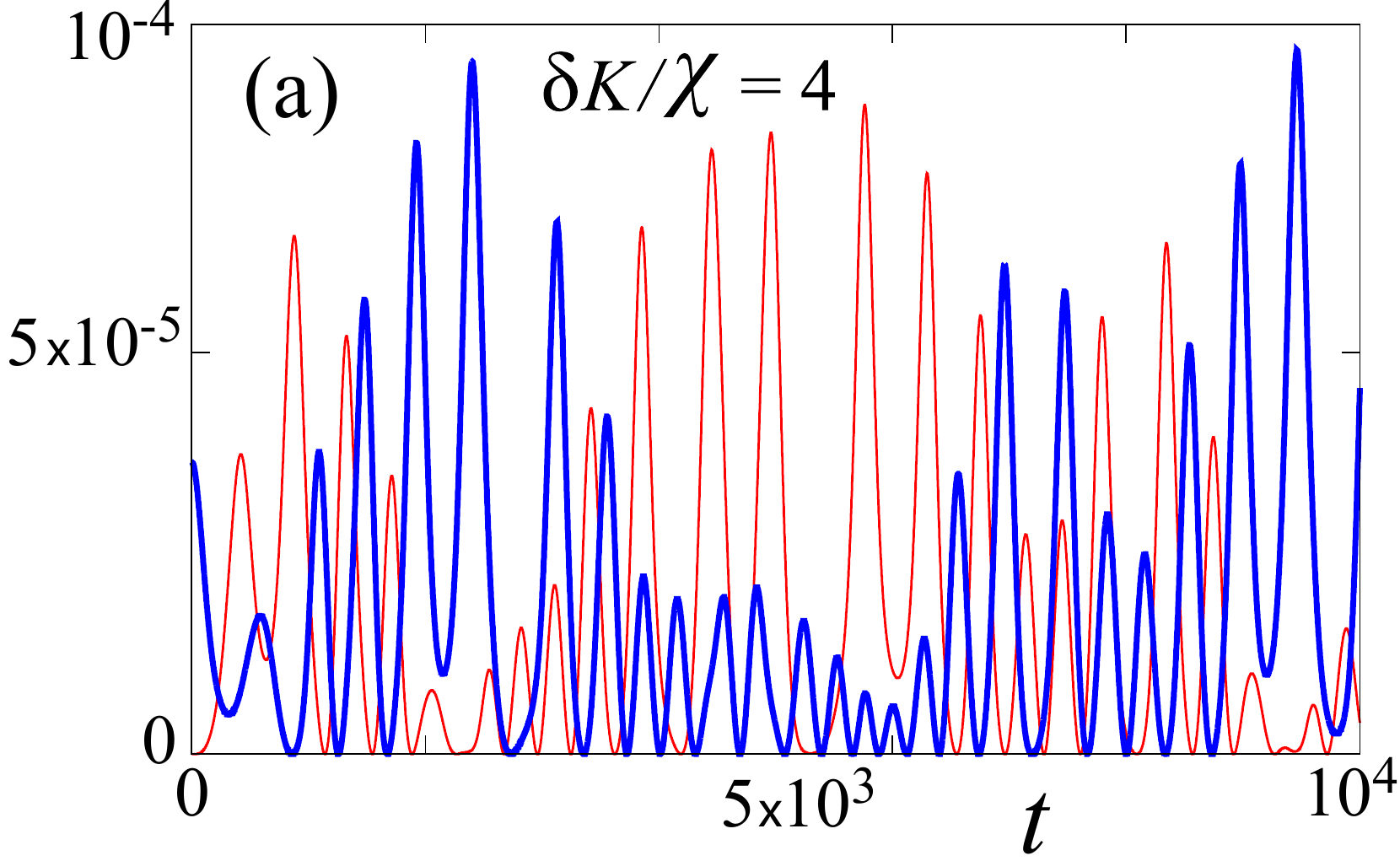}
\includegraphics[scale=0.36]{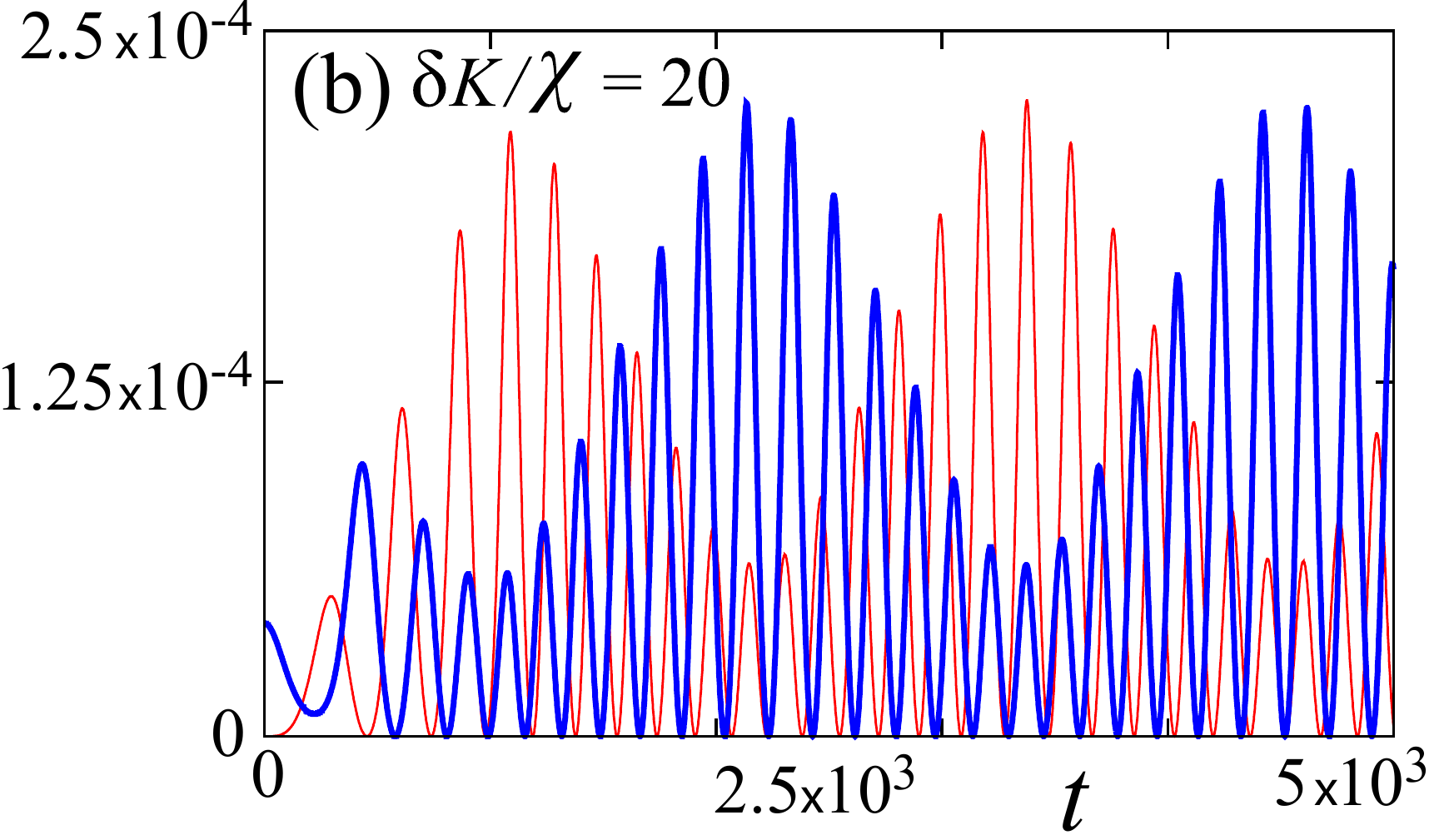}
\includegraphics[scale=0.36]{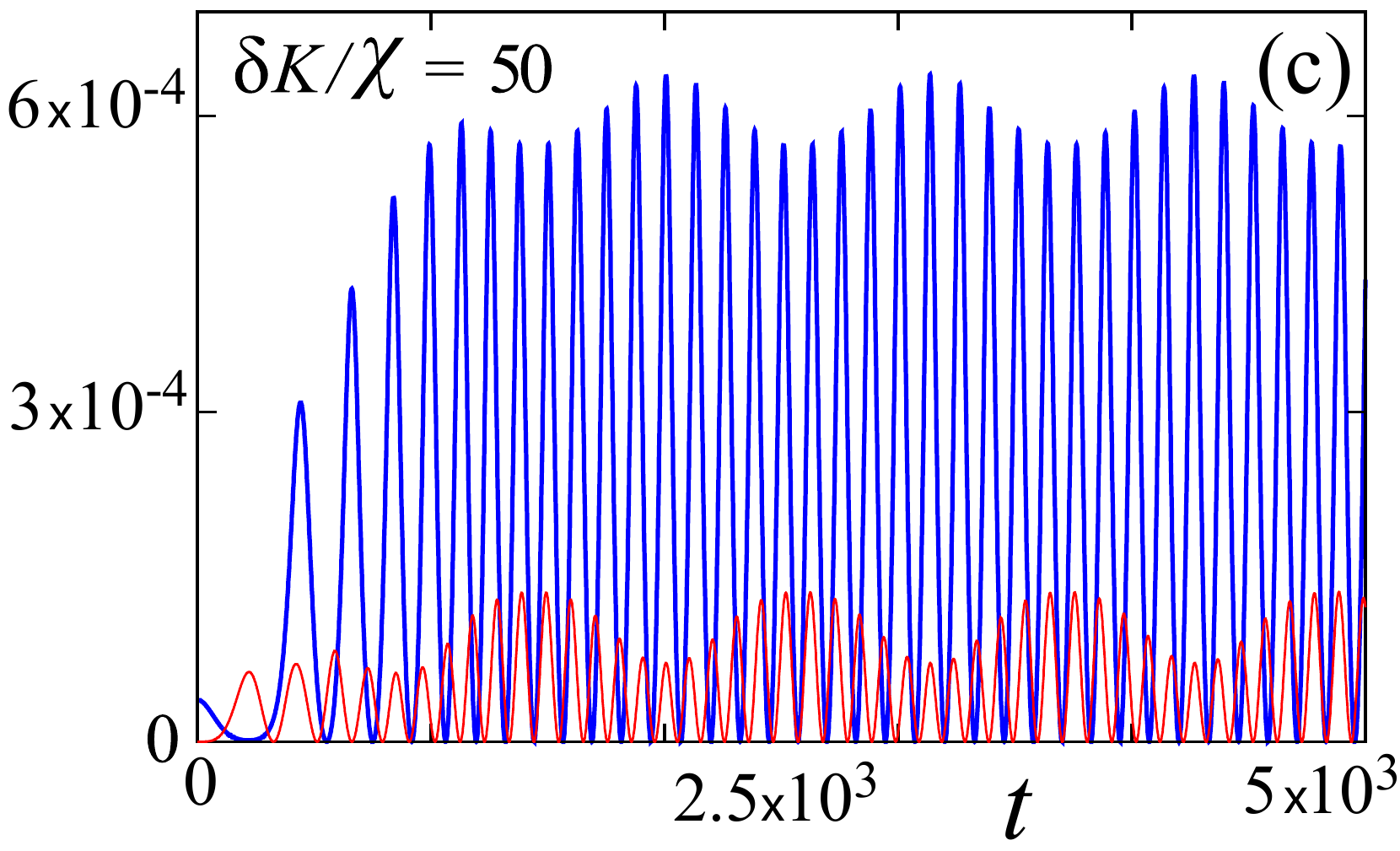}
\caption{Time evolution of the kinetic energies $P_1^2(t)$ (thick blue solid line)
and $P_2^2(t)$ (thin red solid line) 
of the $\tau$ orbit for 
$\delta {\cal K}=4\chi$ and (a) $\beta=4\chi$, (b) $\beta=20\chi$ and (c) $\beta=50\chi$. 
The dipole-dipole interaction parameter is to $\chi=10^{-5}$.}
\label{fi:time_evolution}
\end{figure*} 

For $\beta=5\chi$,  the system shows a sensitive dependence on the electric
field parameter, cf.~\autoref{fi:evolution2}, and
the Poincar\'e surfaces of section exhibit a chaotic sea. A single trajectory with initial conditions 
in this sea covers 
randomly a large portion of the Poincar\'e map, see~\autoref{fi:sos4}a and~\autoref{fi:sos7}a.
In particular, the chaotic sea of these surfaces of section results in strongly fluctuating kinetic energies 
$P_1^2(t)$ and  $P_2^2(t)$, as it is shown in~\autoref{fi:time_evolution}a  for the  orbit $\tau$ with
 $\delta {\cal K}=4\chi$ and $\beta=5\chi$.
For stronger electric fields,  the phase space 
of the system is made up of three different types of regular KAM tori
organized around two stable periodic orbits, and 
kept apart by a separatrix attached to an unstable periodic orbit,
see~\autoref{fi:sos4} and \autoref{fi:sos7}.
Each type of KAM tori corresponds to one of the 
kinetic energy partition regimes detected in~\autoref{fi:evolution2}. Indeed, when the
dipoles are in the energy equipartition regime, the 
trajectory $\tau$ falls inside the KAM torus centered around the stable periodic orbit located on the right hand
side of the Poincar\'e surfaces of section. 
This is  observed for the surfaces of section for 
 $\beta=20\chi$ with $\delta {\cal K}=4\chi$ and $\delta {\cal K}=7\chi$ in~\autoref{fi:sos4}b and~\autoref{fi:sos7}b,
 respectively, and $\beta=50\chi$ and $\delta {\cal K}=7\chi$ in~\autoref{fi:sos7}c. The 
equipartition energy is manifest  in the time evolution of the kinetic energies $P_1^2(t)$
and $P_2^2(t)$ presented in~\autoref{fi:time_evolution}b for $\delta {\cal K}=20\chi$ and $\beta=20\chi$.
In contrast, if the system falls out of the equipartition regime with the first dipole having most of the kinetic energy, 
 the reference trajectory $\tau$ appears in the corresponding Poincar\'e maps inside a different type of KAM torus located at the periphery of the Poincar\'e map, as it is observed for
 $\beta=50\chi$ and $\delta {\cal K}=4\chi$ in~\autoref{fi:sos4}c,
 and for    $\beta=100\chi$ with $\delta {\cal K}=4\chi$  and $\delta {\cal K}=7\chi$
 in~\autoref{fi:sos4}d and~\autoref{fi:sos7}d,  respectively. 
In these orbits, the kinetic energy $P_1^2(t)$ reaches significantly larger values than  $P_2^2(t)$, see for instance, 
$P_1^2(t)$ and $P_2^2(t)$ shown in~\autoref{fi:time_evolution}c for $\delta {\cal K}=4\chi$ and $\beta=50\chi$.
For other values of the excess energy $\delta {\cal K}$, not shown in~\autoref{fi:evolution2}, the second dipole
could have most of the kinetic energy and
 the corresponding Poincar\'e surface of section of the trajectory
 $\tau$ is a  KAM torus located around the central 
 stable periodic orbit.

\section{Conclusions}
\label{sec:conclusions}
We have explored the classical phase space and related
energy transfer mechanisms between two dipoles in the presence of an homogenous
electric field.
The dipoles are described by the  rigid rotor  approximation and are assumed to be fixed in space.
In our numerical study, initially the molecules are at rest in the
stable lowest energy head-tail configuration. At $t=0$,  the system is pushed out of equilibrium by injecting 
a certain amount of kinetic energy  to one of the dipoles.
The following dynamics is investigated by analyzing in particular the  kinetic energies of the dipoles and
their time-averages. 

In the field-free case, and depending on the amount of excess energy in one of the dipoles, 
the system  falls to either an energy equipartition regime or
a non-equipartition one. The transition between these two regimes is abrupt and takes place at
 $\delta {\cal K}=6\chi$.  The
analysis of the phase space structure of the system by means of
Poincar\'e surfaces of section as well as a rotation of the  Hamiltonian provide
the explanation of  this sharp transition.

The impact of the electric field on the energy transfer between the dipoles is quite dramatic.
Depending on the field strength, the system shows  different behaviors where
equipartition, non-equipartition and even chaotic regimes are possible.
If the strengths of the
dipole-dipole and electric field interactions are
comparable, the energy transfer is a chaotic  process and 
the time-averaged kinetic energies strongly
depend on the field parameter  and show rapid and sudden changes.
Again, the  phase space structure of the system by means of the Poincar\'e
surfaces of section provides a global picture of the energy exchange mechanism.

We have here been focusing on an invariant subspace of the full dynamics and, therefore,
it would be a natural continuation of this work to investigate the exchange of energy in
the remaining part of the energy shell. Besides this, an extension
of the system to a linear chain of dipoles is of immediate interest.

\begin{acknowledgments}

R.G.F. acknowledges financial support by the Spanish project FIS2014-54497-P (MINECO)
and the Andalusian research group FQM-207.
M.I. and J.P.S. acknowledge financial support by the Spanish project
MTM-2014-59433-C2-2-P (MINECO).

\end{acknowledgments}

{}
\end{document}